\documentstyle{article}
\title{Decay of metastable phase on
heterogeneous centers with continuous activity}

\author{V.Kurasov}

\date{Victor.Kurasov@pobox.spbu.ru}

\begin{document}
\maketitle
\begin{abstract}

A system of a metastable phase with a pseudo continuous set
of the heterogeneous centers is
considered.
An analytical theory for kinetics
of  decay of the metastable phase  in such a system is constructed.
The free energy of formation of a critical embryo
 is assumed to be known in the macroscopic
 approach as well as the solvatation energy.
The theory constructed below is
based on the quasistationary approximation for the nucleation rate.
\end{abstract}

\pagebreak

\section{Introduction}

The theory considered here will be based on the macroscopic approximation of
the height of the activation barrier.
This approximation lies in the base of the classical theory of nucleation.
The theory presented below forms the part in the program announced in
\cite{aero1}.
All bibliographical remarks can be found in \cite{aero1}.

The further investigation of the heterogeneous condensation requires
to take into consideration the process of the heterogeneous decay on
the band of centers with the different activity. Speaking about
the activity
of the heterogeneous centers we mean that the different activity initiates
the different height of the activation barrier $\Delta F$, i.e. the
difference between the free energy of the critical embryo $F_{c}$ and
the free energy of solvatation $G$. The set
of the different activities of the heterogeneous centers is
so dense that we can regard it as a pseudo continuous
one. In the atmosphere this fact is
proved experimentally. In the case of the "solid nucleus of
condensation with the weak interaction"
the continuous size of the nucleus ensures the continuous set of $\Delta F$.

Nevertheless the theory of the heterogeneous decay was constructed only
for one type of the heterogeneous centers. So, the task to construct the
 kinetic theory of
the decay for the system with the continuous set of
the heterogeneous centers is rather
essential. It will be completely fulfilled here.

We shall use the ordinary physical assumptions formulated in \cite{aero1}
including the following ones:
\begin{itemize}
\item
the thermodynamic description of the critical embryo,
\item
the random homogeneous space distribution of the
heterogeneous centers,
\item
the free-molecular regime of the droplets growth,
\item
the homogeneous
external conditions for the   temperature and for the pressure,
\item
rather a high
activation barrier for the nucleus which are really going to be the centers
of the droplets and aren't absolutely exhausted in the process.
\end{itemize}
As far as the most interesting characteristics of this process are the
 numbers of the heterogeneously
formed droplets of the different activities
we shall estimate the accuracy of the theory
by
 the error of the obtained solutions for
 these values. The whole process of the metastable decay
can be split into the two periods: the
period of the  essential  formation of the droplets and
the period of the essential consumption of the metastable
phase. At first we shall
investigate the period of the the droplets formation.
The unit volume is considered.

The characteristic moment $t_*$ of the process of formation can be chosen
as the initial moment. The values at the moment $t_*$ will be marked by
the lower idex "*".

We assume the total number of the heterogeneous centers to be constant in
time.

We shall define the activity of the heterogeneous center as some parameter $w$
which
is proportional to the height of the activation barrier
\begin{equation} \label{**}
\Delta F (w) = \Delta F \mid_{w=0} - \lambda w
\end{equation}
with some positive parameter $\lambda$.

The total number of the heterogeneous centers
with the given activity $w$ will be marked by $\eta_{tot}(w)$. Naturally
$\eta_{tot}(w)$ is rather  a smooth function of $w$.

          The density of the molecules in the
equilibrium vapor is marked
by \( n_{\infty} \), the
density of the molecules in the real vapor  is marked by \( n\).
 The power of the metastability will be characterized by the
value of the supersaturation
$$ \zeta = \frac{ n - n_{\infty} }{ n_{\infty} } $$
We shall define the super-critical embryos as the "droplets".
Every droplet is described by the number of the molecules
\( \nu \)  or by the linear size \( \rho = \nu^{1/3} \) .
Due to the  free-molecular regime of vapor consumption  we have
$$ \frac{d\rho}{dt} = \zeta \alpha \tau^{-1} $$
where \( \alpha \) is the condensation coefficient and \( \tau \) is
some characteristic
time between collisions obtained from the gas kinetic theory.

The frontal type of the size spectrum
allows to introduce the frontal size \( z \)
according to
\begin{equation}
\label{2}
z = \int_{0}^{t} \zeta \alpha \tau^{-1} dt'
\end{equation}
Until the coalescence \cite{15},\cite{16}
which isn't considered here
equation (\ref{2}) ensures the growth of \( z \) in time and can be
inverted
\begin{equation}
t(z) =  \int_{0}^{z} \tau \alpha^{-1} \frac{dz}{\zeta(z)}
\end{equation}
Hence, all values dependent on time become the values dependent on
\( z\) and the relative size \( x=z-\rho \) can be introduced.
During the whole evolution the droplet has one and the same
value of the variable \( x \).
Consider \( t(x) \) as the moment when the droplet with the given $x$ has
been formed (as a droplet). We can see that the functions of time are
the  functions
of \( x \) .
Hence, we can see that the kinetic equation can be reduced to the
fact that every droplet keeps the constant value of $x$. To reconstruct the
picture
of the evolution one must
establish the dependencies $t(z)$ and $\zeta(x)$.

\section{Condensation equations system}

We shall mark by the argument \( \infty \) the total values
formed during the whole condensation process.

We must take into account the reduction of $\zeta(0)$ to some value
$\Phi_{*}$ due to the consumption of the vapor molecules in the process of
solvatation.
For $\Phi_{*}$ we have
\begin{equation}
\Phi_{*} = \zeta (0) - \frac{\int dw \eta_{tot}(w) \nu_{e}(w)}{n_{\infty}}
\end{equation}
where $\nu_{e}$ is  the  number  of  the molecules  of  the  condensated
substance in
the equilibrium embryo.
The discrepancy between $\Phi_{*} $ and $\zeta(0)$
for the monotonic interaction between the molecules in the
liquid phase  and the
heterogeneous center can be estimated as
\begin{equation}
\Phi_{*} - \zeta(0) \geq \eta^{tot} \frac{\nu_{c\ hom}}{n_{\infty}}
\end{equation}
Here $\eta^{tot}$ is the total number of the heterogeneous centers of all
types:
\begin{equation}
\eta^{tot}= \int dw \eta_{tot}(w)
\end{equation}

The following statements are valid in the  further consideration:

(1) The main role in the vapor consumption during the evolution
is played by the super-critical embryos, i.e. by the droplets.

(2) The quasistationary approximation for the nucleation rate is valid during
the
period of the essential formation of the droplets.

They are proved analytically.

The justification of the second statement uses the estimate for the times
 \( t^{s}_{i} \)
of the establishing of the stationary
state in the  near-critical region which can be found in \cite{3},
\cite{17}
(for heterogeneous barrier  the consideration is  quite analogous).

Let \( f_{s} \) be the stationary value of
the size distribution for the heterogeneously  formed droplets
 measured in the units of \( n_{\infty} \).
It can be presented in the following form
\begin{equation}
f_{s} = f_{\zeta}(\zeta (x),w) \eta(x,w)
\end{equation}
where $\eta(x,w)$ is the number of the heterogeneous centers
free from the super-critical embryos and $f_{\zeta}$ is given by the following
formula \cite{18}
\begin{equation}
f_{\zeta}=\frac
{
 W^{+}_{c} \exp(-\Delta F(\zeta,w)) \tau
}
{
 \pi^{1/2} \Delta_{e} \nu  \Delta_{c} \nu  \zeta \alpha  n_{\infty}
}
\end{equation}
where $W^{+}$ is the number of the molecules interacting with the droplet in
the unit
of time, $\Delta_{e} \nu$  is the width of the equilibrium distribution
$$
\Delta_{e} \nu = \sum_{\nu=1}^{\nu=(\nu_{c}+\nu_{e})/2}\exp(-F(\nu)+G)
$$
and $\Delta_{c} \nu$ is the halfwidth of the near-critical region
$$
\Delta_{c} \nu =
\frac{2^{1/2}}{\mid
(\frac{\partial^{2} F }{\partial \nu^{2}})_{\nu=\nu_{c}} \mid^{1/2} }
$$
Index "$c$" marks the values for the critical embryo.
Certainly, $W^+$, $\Delta_c \nu$ and $\Delta_e \nu$ are some smooth functions
of $w$ and we neglect this dependence.

We shall mark by \( n_{\infty} g(w) \) the total number of the vapor
molecules in the heterogeneously formed   droplets
 on the centers of the activity $w$.
To simplify the formulas we shall use $$ \theta(w) =
{\eta(w)}/{\eta_{tot}(w) } $$

 We obtain for \( g,  \theta \) the following
equations
\begin{equation}
g(z,w) =   \int_{0}^{z} (z-x)^{3} f_{\zeta}(\zeta(x),w) \eta(x,w) dx
\end{equation}
\begin{equation}
\theta(z,w) = \exp ( - n_{\infty} \int_{0}^{z} f_{\zeta}(\zeta(x),w) dx )
\end{equation}

As far as we measure the accuracy of the theory in terms of the error
in the droplets number
 we  define these values as the following ones:
\begin{equation}
N(z,w) = \eta_{tot}(w) ( 1 - \theta(z,w))
\end{equation}
The total number of the droplets is
\begin{equation}
N^{tot} = \int \eta_{tot}(w)( 1 - \theta(z,w)) dw
 = \int N(z,w) dw
\end{equation}

For the majority of the types of the heterogeneous centers
 the following approximations  for the  nucleation rates
 are valid during the period of the essential  formation of the droplets
\begin{equation}
\label{7}
f_{\zeta}(\zeta(x),w) =
f_{\zeta}(\Phi_{*},w) \mid_{w=0}
\exp ( \Gamma \frac{ ( \zeta - \Phi_{*} ) }
{ \Phi_{*} } )
 \exp( w \lambda )
\end{equation}
where
\begin{equation} \label{8}
\Gamma = -\Phi_{*}
\frac{d \Delta_{i} F(\zeta)}{d \zeta }  \mid_{\zeta=\Phi_{*}}
\end{equation}
and \( \Delta F \) is the height of the heterogeneous activation barrier.
The validity of these approximations can be analytically justified
for the heterogeneous embryos with
the interaction between the center and the molecules of the condensated phase
 weaker or equal than the function reciprocal to a space distance.
Then  we can imagine the center as the hard sphere with a weak interaction  on
which
the embryo  is formed.

The dependence of $\Gamma$ on $w$ is rather a weak one. So we can put
\begin{equation}
\Gamma(w) = \Gamma \mid_{w=0}
\end{equation}
for some essential part of the activity spectrum.

Using the conservation laws for the heterogeneous centers
and for the molecules of the condensated
substance   we obtain for \( g,  \theta \) the following
equations
\begin{equation}
g(z,w) = f_{*}  \int_{0}^{z} (z-x)^{3}
\exp ( -\Gamma \frac{ g^{tot}  }
{ \Phi_{*} } )
\theta dx \eta_{tot}(w) \exp( w \lambda )
\equiv
G_w (g^{tot}, \theta)
\end{equation}
\begin{equation}\label{17}
g^{tot} = \int dw g(z,w)
\end{equation}
\begin{equation}\label{18}
\theta(z,w)
 = \exp ( - f_{*}\exp(  \lambda w)  n_{\infty}  \int_{0}^{z}
\exp (  - \Gamma \frac{ g^{tot}  }
{ \Phi_{*} } ) dx )
\equiv
S_w ( g^{tot})
\end{equation}
where $f_{*} = f_{\zeta}(\Phi_{*},w=0)
$

These equations form the closed system of the equations for the condensation
kinetics. This system will be the subject of our investigation.

The size spectrum  can be found as the following one
\begin{equation}
f(x,w) = f_{*} \exp(\lambda w)
\exp ( -\Gamma \frac{ \int  dw g(x,w)  }
{ \Phi_{*} } )
\theta(x,w)
\eta_{tot} (w)
\end{equation}

\section{Iteration procedure}
The systems
like (\ref{17})-(\ref{18}) are the ordinary solved by the iteration procedure.
 It can be constructed by the following way: For the initial approximations
we choose:
\begin{equation}
g_{0}(z,w) = 0 \ \ \ \ \theta_{0} = 1
\end{equation}
Ordinary the recurrent procedure is defined according to
\begin{equation}
g_{i+1}(z,w) = G_w (g^{tot}_{i}, \theta_{i}(w) )
\end{equation}
\begin{equation}
g^{tot}_{i}(z) = \int dw g_{i}(z,w)
\end{equation}
\begin{equation} \label{***}
\theta_{i+1}(z,w) = S_w(g^{tot}_{i})
\end{equation}
The remarkable monotonic properties of $G$, $\int dw$ and $S$ ensures
the guaranties for the error of the approximation. In this situation they can
be observed
also in the manner analogous to the case of ordinary heterogeneous
condensation.
 Certainly, for the calculation of the iterations
 we must use some expression for $\eta_{tot}(w)$. So at first
we shall consider the limit cases and obtain some estimates.

An interesting fact goes from the absence of $\eta_{tot}$ in the r.h.s. of
(\ref{***}). So for some rather an arbitrary $\eta_{tot}$ the power of exhaustion
will be determined  only by $w$. The centers with the high activity $w$
are almost exhausted during the process of condensation. The centers with
the relatively small activity $w$ remain unexhausted. The intermediate region
has
rather a
small size. To give the qualitative estimates let us notice that the
supersaturation
$\zeta$ appears only in the function $f_{\zeta}$. For the behavior of
$f_{\zeta}$
we can obtain the following
estimate
\begin{equation} \label{888}
f_{\zeta} \sim f_* \exp(-const x^{\epsilon}) \ \ \ \ \  3\le \epsilon \le 4
\end{equation}
This estimate  goes from the general obvious fact that the spectrum of the
droplets
must be  wider than the monodisperse one and the intensity of formation
must
decrease in the $x$ scale. So we see that the width $\Delta x$ of the spectrum
 is a well defined value. We must stress that (\ref{888}) is valid not only
for the separate process of condensation on the centers with some separate
activity  but also for condensation on the centers with
the set of all activities.

The most rough approximation for the exhaustion of the heterogeneous
centers  can be obtained
from the first approximation in the  iteration procedure. We have
\begin{equation}
\theta = \exp [ -  f_{*} \exp(\lambda w) n_{\infty} z ]
\end{equation}
and for the final value
\begin{equation}
\theta_{final} = \exp [ - f_{*} \exp(\lambda w) n_{\infty} \Delta x ]
\end{equation}
Let us define $w_{<}$ and $w_{>}$ according to the following expressions
\begin{equation}
w_{<} = w_{0} - \frac{\epsilon}{\lambda}
\end{equation}
\begin{equation}
w_{>} = w_{0} + \frac{\epsilon}{\lambda}
\end{equation}
where
\begin{equation}
w_{0} = \frac{1}{\lambda} \ln (\frac{ \ln 2 }{ f_{*} n_{\infty} \Delta x } )
\end{equation}
$$
\epsilon \sim 1
$$
For $w > w_{>}$ almost all heterogeneous centers are exhausted. For $w<w_{<}$
almost all heterogeneous centers remain free.
This remark  allows to rewrite
 the expression for $g^{tot}$ in the following form
\begin{equation}
g^{tot} = \int_{w_{<}}^{w_{>}} dw g(z,w)
+ \int_{w_>}^{\infty} \eta_{tot}(w) dw \frac{z^3}{n_{\infty}}
\end{equation}
The size of the intermediate region has in the scale of $w$
of the order of $1/\lambda$. This size
corresponds
to the variation of $\Delta F$ of the order of $1$. As far as $\Delta F$ has
rather big values, the relative variation of $\Delta F$ in the intermediate
region
is
rather small. So we can put in this region
\begin{equation}
\eta_{tot}(w) = \eta_{*} = const
\end{equation}
We can spread this approximation over the region $w>w_{>}$ because the real
activity of these centers isn't important. Certainly, we must obtain the
boundary $w_{max}$ of this region from the following equation:
\begin{equation}
\int_{w_{>}}^{w_{max}} \eta_{*} dw = \eta_{*}(w_{max}-w_{>}) =
\int_{w_{>}}^{\infty} \eta_{tot}(w) dw
\end{equation}
For $w<w_{<}$ the accuracy of the approximation is
not so essential because these centers remain
free.

After the scale transformations $\Gamma g^{tot}/\Phi_{*} \rightarrow G ; \ \ \
\Gamma g /\Phi_{*} \rightarrow g$ we have
\begin{equation} \label{a1}
g=\frac{\Gamma f_{*} \eta_{*}} {\Phi_{*}} \exp(\lambda w) \int_{0}^{z}
(z-x)^3 \exp(-G) \theta dx
\end{equation}
\begin{equation} \label{a2}
G = \int_{-\infty}^{w_{max}} dw g
\end{equation}
\begin{equation} \label{a3}
\theta = \exp [ - f_{*} n_{\infty} \exp(\lambda w) \int_{0}^{z} \exp(-G) dx ]
\end{equation}
By the appropriate shift of $w$ one can put $w_{max}$ to the zero. By the
appropriate
 scale of $w$ one can put $\lambda = 1$.  By the appropriate choice of $z$
we put the coefficient in (\ref{a1}) to unity.  As a result we have only one
parameter - the coefficient in the last equation. We shall mark it by "$A$".

The final form of the system of the balance equations is the following one
\begin{equation} \label{a4}
g=\exp(w) \int_{0}^{z} (z-x)^3 \exp(-G) \theta dx
\end{equation}
\begin{equation} \label{a5}
G = \int_{-\infty}^{0} dw g
\end{equation}
\begin{equation} \label{a6}
\theta = \exp[-A \exp(w) \int_{0}^{z} \exp(-G) dx ]
\end{equation}
An iteration procedure can be constructed by the following way
\begin{equation}
g_{i+1} = \exp(w) \int_0^z (z-x)^3 \exp(-G_i) \theta_i dx
\end{equation}
\begin{equation}
G_i = \int_{-\infty}^0 dw g_i
\end{equation}
\begin{equation}
\theta_{i+1} = \exp(-A \exp(w) \int_0^z \exp(-G_i) dx)
\end{equation}
\begin{equation}
g_0 = 0 \ \ \ \  \theta_0 = 1
\end{equation}
So, we have in the first iteration
\begin{equation}
g_1 = \exp(w) \frac{z^4}{4}
\end{equation}
\begin{equation}
G_1 = \int_{-\infty}^0 dw g_1 = \frac{z^4}{4}
\end{equation}
\begin{equation}
\theta_1 = \exp( - A \exp(w) z)
\end{equation}
The second approximation gives for $\theta$ the following result
\begin{equation}
\theta_{2} = \exp[-A \exp(w) \int_{0}^{z} \exp(-\frac{z^4}{4}) dz]
\end{equation}
and for  the final value
\begin{equation}
\theta_{final\ 2}= \exp[-A \exp(w)  1.28^{1/4}]
\end{equation}
Then for $N^{tot}$ we have
\begin{equation}
N^{tot} = \int_{-\infty}^{0}[1-\exp(-A\exp(w) 1.28^{1/4})] dw    \eta_*
\end{equation}
For $g$ in the second approximation we have
\begin{equation}
g_{2} = \exp(w) \int_{0}^{z} (z-x)^3 \exp(-A \exp(w) x ) \exp(- x^4/4) dx
\end{equation}
The analytical expression for these and the further iterations can not be
obtained.
So we shall investigate the reduced iterations. In this procedure
$\theta_i$ is substituted by $\theta_{i+1}$ in the expression for $g_{i+1}$.
 So, we have
\begin{equation}
\theta_{1} = \exp[-A \exp(w) z]
\end{equation}
\begin{equation}
g_{1} = \exp(w) \int_{0}^{z} (z-x)^3 \exp(-A \exp(w) x) dx
\end{equation}
\begin{equation}
g_{1} =
\frac{1}{A}z^3
- \frac{3}{A^2}z^2 \exp(-w)
+\frac{6}{A^3}z \exp(-2w)
- \frac{6}{A^4} \exp(-3w)
+ \exp(-3w) 6 A^{-4} \exp(-A \exp(w) z)
\end{equation}
\begin{equation}
G_{1} = \int_{0}^{z} (z-x)^3 \int_{-\infty}^{0} \exp(w) \exp(- A \exp(w) x) dx dw
\end{equation}
\begin{equation} \label{r47}
G_{1}
=\int_{0}^{z} (z-x)^3 \frac{1}{Ax} [1-\exp(-Ax)] dx
\end{equation}
After the decomposition of the last subintegral expression we have
\begin{equation}
G_{1} = z^3  I_{0} - 3 z^2 \frac{I_{1}}{A} + 3 z  \frac{I_{2}}{A}
-  \frac{I_{3}}{A}
\end{equation}
where
\begin{equation}
I_{0} = \int_{0}^{z} \frac{1}{Ax} [1-\exp(-Ax)] dx
\end{equation}
\begin{equation} \label{i1}
I_{1} = z - \frac{1- \exp(-Az)}{A}
\end{equation}
\begin{equation}\label{i2}
I_{2} = \frac{z^2}{2} + \frac{1}{A^2} Az \exp(-Az) -
\frac{1}{A^2} (1-\exp(-Az))
\end{equation}
\begin{equation}\label{i3}
I_{3} = \frac{z^3}{3} + \frac{1}{A^3} (Az)^2 \exp(-Az)
+ \frac{2}{A^3} Az \exp(-Az)
- \frac{2}{A^3}(1-\exp(-Az))
\end{equation}
For $I_{0}$ we have the following result
%% tom 1 str 139
\begin{equation} \label{c13}
\int_{0}^{z} \frac{1-\exp(-y)}{y} dy = E_{1}(z) + \gamma +\ln (z)
\end{equation}
where
\begin{equation}
\gamma = 0.57721\ \ \ \ E_{1}(z) = - Ei(z)
\end{equation}
and $Ei$ is  the exponential integral function. We can present this result in
the following approximate form
%% Luc str 115
\begin{equation} \label{c14}
\int_{0}^{z} \frac{1-\exp(-y)}{y} dy \approx z \frac{
\frac{17}{3} z^2 + 60 z +480
}{
z^3 +24 z^2 +180 z +480
}
\end{equation}
The final  approximation for $\theta$ has the following form
\begin{equation} \label{b1}
\theta_{2} = \exp\{ - A \exp(w) \int_{0}^{z} \exp(- G_{1} ) dx \}
\end{equation}
This expression can be calculated only approximately. Notice that $G$
increases
faster than $x^3$. The action of the subintegral factors like the functions
$\exp(-const x^n)$  approximately
leads simply to the cut-off of the region of
integration                      when $n \geq 3$
\begin{equation} \label{v1}
\exp(-const x^n) \approx \Theta(const^{-1/n} - x )
\end{equation}
In order to obtain the position of the end of the process of formation
 we must solve
the algebraic equation
\begin{equation}
G_{1} = 1
\end{equation}
We can also use the other method.
From (\ref{r47}) after the decomposition of the exponent one can get the
series
containing some non-negative powers of $A$. So, for small $A$ one can
restrict the series by several first terms and promote the calculations by
the following method.
All integrals with the combination of $\exp(-const x)$ and
$\exp(-const x^2)$ can be calculated with the help of the Boyd's approximation
\begin{equation}
\frac{\pi/2}{(z^2 + \pi)^{1/2} + (\pi -1)z}
\leq
\exp(z^2)
\int_{z}^{\infty} \exp(-t^2) dt
\leq
\frac{\pi/2}{((\pi-2)^2 z^2 +\pi)^{1/2} +2 z}
\end{equation}
All terms like $\exp(-const x^n) \ \ \
n = 3,4,...$  can be treated
as some functions like the Heavisaid's one.

The value of $\theta_{2}$ is the base for the final approximation
$N_{tot}(w)$.
The total number of the droplets can be found directly from (\ref{b1}) after
the integration
\begin{equation} \label{c10}
N^{tot} = \eta_{*} \int_{-\infty}^{0} [1-\theta (z,w) dw]
\end{equation}
In the second approximation we have
\begin{equation} \label{c11}
N^{tot} = \eta_{*} \int_{-\infty}^{0} [1- \exp(-B \exp(-w)) dw]
\end{equation}
where
\begin{equation}
B = A \int_{0}^{z} \exp(-G_{1}) dx
\end{equation}
is calculated earlier. Hence, for the final expression we obtain the integral
which is already calculated in (\ref{c13}), (\ref{c14}).

The alternative method is the following one. At least after the remark
 that in
the essential
region there is $\exp(-G_1) \leq 1$ we can decompose this exponent as a function of
$G_1$ and get the answer for
$\int_0^z \exp(-G_1) dx$ in the polynomial form.

\section{Universal solution method}

System of equations (\ref{a4}) - (\ref{a6}) doesn't allow the universal
solution as in the case of the homogeneous condensation. Equation (\ref{a4})
hasn't $\theta$ in the r.s.h.. But the operator in the r.h.s. of (\ref{a4})
ensures
rather a rapid convergence of the iterations. The worst situation for
the iteration procedure is when $A=0$. In this situation we have the universal
system
\begin{equation}
g = \exp(w) \int_{0}^{z} (z-x)^3 \exp(-G) dx
\end{equation}
\begin{equation}
G=\int_{-\infty}^{0} g dw
\end{equation}
This system has the universal solution which will be marked by $G_{0}$.

This universal form of the size spectrum is shown by Figure 1.

The condensation process is determined  by the first three momentums of the
distribution
function
\begin{equation} \label{cx0}
\mu_{i}(z) = \int_{0}^{z} x^{i} \exp(-G) \theta dx
\end{equation}
and by the zero momentum
\begin{equation} \label{cx1}
\mu_{+}(z) = \int_{0}^{z}  \exp(-G) \theta dx
\end{equation}
In the pseudo homogeneous case it is determined by
\begin{equation} \label{cx0dd}
\mu_{i}(z) = \int_{0}^{z} x^{i} \exp(-G) dx
\end{equation}
\begin{equation} \label{cx1dd}
\mu_{+}(z) = \int_{0}^{z}  \exp(-G)  dx
\end{equation}

After the end of rather a short period of the intensive formation we can
substitute
in (\ref{cx0}) and (\ref{cx1}) $\infty$ instead of $z$ in the region of
the integration. So the further evolution will be determined  by
the first four momentums
$\mu_{i}(\infty)$\footnote{
The main role is played by $\mu_{+}$.}. As far as for $G$ we substitute  the
universal
solution $G_{0}$
(when $A=0$), the values of $\mu_{i}$ are the universal constants which can
be obtained by the unique numerical solution of the last system.

Now we return to the previous iteration procedure (with the new initial
approximation).
We can use $G_{0}$ as the initial approximation in the iteration procedure.
So we have
\begin{equation} \label{d1}
\theta_{1} = \exp[-A \exp(w) \int_{0}^{z} \exp(-G_{0}) dx]
\end{equation}
\begin{equation}
g_{2} = \exp(w) \int_{0}^{z} (z-x)^3 \exp(-G_{0}) \exp(-A \exp(w) \int_{0}^{x}
\exp(-G_{0}) dx') dx
\end{equation}
\begin{equation}
G_{2} = \int_{-\infty}^{0}
\exp(w) \int_{0}^{z} (z-x)^3 \exp(-G_{0}) \exp(-A \exp(w) \int_{0}^{x}
\exp(-G_{0}) dx') dx
dw
\end{equation}
The decomposition of the exponents gives
\begin{equation}
G_2 = \sum_{i=0}^{\infty} \frac{(-A)^i}{(i+1)!} P_i(z)
\end{equation}
where
\begin{equation}
P_{i} = \int_{0}^{z} (z-x)^3 \exp(-G_{0}) I_{00}^{i}(x) dx
\end{equation}
\begin{equation}
I_{00}(x) = \int_{0}^{x} \exp(-G_{0})dx
\end{equation}
Then the expression for the function $B(z=\infty)$ will be the following
one
\begin{equation}
B(z=\infty) = A \int_{0}^{\infty} \exp(-G_2) dx =
A \int_0^{\infty}
\exp(-P_0(z))
\exp(-\sum_{i=1}^{\infty} \frac{(-A)^i}{(i+1)!} P_i(z)) dz
\end{equation}
The decomposition of the last  exponent leads to
\begin{equation}
B(z=\infty) = \prod_{i=1}^{\infty} \sum_{j=0}^{\infty} \frac{(-1)^j}{j!} A
\frac{(-A)^{ij}}{((i+1)!)^j} C_{ij}
\end{equation}
where
\begin{equation}
C_{ij} =
\int_{0}^{\infty} \exp(-P_{0}(x)) P_{i}^{j}  dx
\end{equation}
are the universal constants.
For the total number of the droplets with the given activity we have
\begin{equation} \label{d9}
N_{tot}(w) = \eta_{*} (1 - \exp(-B(z=\infty) \exp(w)))
\end{equation}
For $N^{tot}$ we repeat  expressions (\ref{c10}), (\ref{c11}), (\ref{c13})
with the new value of $B$ and obtain the final result.

In the same manner one can take into account the deviation of $\eta_{*}$
from the constant value.
After the decomposition of $\eta_{*}$ into the Tailor's series we substitute
the
initial approximation $g_{0}$ and act in the manner similar to (\ref{d1}) -
(\ref{d9}). As a result we have some expansions in the powers of the parameter
$A$,
the  derivatives  of $\eta_{tot}$ and the universal constants.

There is another possibility to observe the universal solution in  the
wide
class of the possible situations. We shall speak about the case of the
"developed
spectrum of activities" in the following situation. Earlier we extracted
in the spectrum of activities the three regions: the region of rather
an active
centers (they are almost exhausted), the region of the centers with the
small
activity (they remain practically free during the whole process) and
the intermediate region. We have noticed that in the majority of the
situations
the intermediate region has a relatively small size in comparison with the
active
region (the unactive region has no size - it can be spread till infinity).
In this case one can fulfill some further simplifications.

Let us transmit the point $w=0$ to the special activity for which
$$
\theta \mid_{w=0} (\infty) = \frac{1}{2}
$$
i.e. the half of the centers became the centers of droplets.
Then the system of the condensation equations
after the rescaling of $w$ can be written as the following one
\begin{equation}
g = a_0 \exp(w) \int_{0}^{z} (z-x)^3 \exp(-G) \theta dx
\end{equation}
\begin{equation}
G = \int_{-\infty}^{w_{+}} g(w)  dw
\end{equation}
\begin{equation}
\theta = \exp( - a_1 \exp(w) \int_{0}^{z} \exp(-G) dx )
\end{equation}
where $w_{+}$ is the upper boundary of the spectrum and $a_0, \ a_1$ are some
constants. After the rescaling of $z,x$ one can put $a_0 = 1$. The condition
of
the choice of $w=0$ gives for $a_1$:
\begin{equation}
a_1 = \frac{\ln 2}{\int_{0}^{\infty} \exp(-G) dx}
\end{equation}
So, there remains only one parameter $w_{+}$.

Note that the result of the process of condensation will lead to the
following functional form for the number of the heterogeneous centers
\begin{equation}
\eta = \eta_{tot} \exp(  -  const \ \exp(w) \int_{0}^{\infty}
\exp(-G) dx)
\end{equation}
after the process of decay
or
\begin{equation}
\eta = \eta_{tot} \exp(  -  const \ \exp(w) \int_{-\infty}^{\infty}
\exp(x -G) dx)
\end{equation}
after condensation under the dynamic conditions
where $\eta_{tot}$ is the real initial (approximately constant) value
of the number of the heterogeneous centers. The value of $\eta_*$ must
be now substituted by $\eta_{tot}$. In any case
as far as the integral gives some constant value the functional form is
the following one
\begin{equation}
\eta = \eta_{tot} \exp(  -  const \ \exp(w))
\end{equation}
or
\begin{equation}
\theta_{init} =  \exp(  -  const \ \exp(w))
\end{equation}

After the observed process of condensation the cut-off of the spectrum
$w_+$
will be unessential when $\theta_{init}(w_+) \rightarrow 0$. This initiates
the more natural definition $w_+ \rightarrow \infty$
of the activity spectrum without rather an artificial parameter $w_+$.

Now we  substitute this functional dependence into the system of the
condensation equations. Then we  fulfill the integration
over $w$ and come to
\begin{equation}
G \sim \int_{0}^{z} \frac{(z-x)^3 \exp(-G)}
{a+\frac{\ln 2}{\int_{0}^{\infty} \exp(-G(x')) dx'} \int_{0}^{z} \exp(-G(x'))
dx' }  dx
\end{equation}
where the coefficient $a$ is initiated by the new form of the
activity spectrum.

For simplicity one can cancel the coefficient
$$
\frac{\ln 2}{\int_{0}^{\infty} \exp(-G(x')) dx'}
$$
As far as one can see that
$
\int_{0}^{\infty} \exp(-G(x')) dx'
$
has the values which are situated
not so far from $1$ the cancellation of this coefficient
leads only to some unsufficient shift of the zero point in the scale of
the activities.

Then we come to
\begin{equation}
G \sim \int_{0}^{z} \frac{(z-x)^3 \exp(-G)}
{a+ \int_{0}^{z} \exp(-G(x'))
dx' }  dx
\end{equation}

The main essential problem is to establish the form of the size spectrum.
Then the system of the condensation equations can be reduced to the algebraic
system.

Now we are going to show that the form of the spectrum has the weak dependence
on the value of the parameter $a$. In the case
$a \rightarrow 0$ the last equation can be reduced to
\begin{equation}
G \sim \int_{0}^{z} \frac{(z-x)^3 \exp(-G)}
{ \int_{0}^{z} \exp(-G(x'))
dx' }  dx
\end{equation}
which resembles the homogeneous case but with the increasing function
 $
 \int_{0}^{z} \exp(-G(x'))
dx'
$ in the denominator.

When $a \rightarrow \infty$ we have the pure homogeneous case:
\begin{equation}
G \sim \int_{0}^{z} (z-x)^3 \exp(-G)  dx
\end{equation}

Recall that in the absence of this function the spectrum $\exp(-G)$ has
rather a simple approximate qualitative behavior:
$$
\exp(-G)                                 \sim 1 \ \ \ z<z_{bound}
$$
$$
\exp(-G)                                 \sim 0 \ \ \ z>z_{bound}
$$
where the parameter $z_{bound}$ corresponds to $G(z_{bound}) = 1$.

Note that this function isn't so sharp function of $z$ as $(z-x)^3$ is.
The behavior of this function is the following
$$
 \int_{0}^{z} \exp(-G(x'))
dx'
\sim z \ \ \ z<z_{bound}
$$
$$
 \int_{0}^{z} \exp(-G(x'))
dx'
\sim const \ \ \ z>z_{bound}
$$
So the qualitative behavior of $G$ will be the same, only the value of
the characteristic length of the spectrum $z_{bound}$ will be another.

In order to get the same length of the spectrum we rescale $z$ in such
a manner that the parameter $a+c$ appears in front of the external integral.

Then we come to the following equation
\begin{equation}
G  = \int_{0}^{z} \frac{(z-x)^3 \exp(-G)}
{\frac{a}{a+c}+ \frac{1}{a+c} \int_{0}^{z} \exp(-G(x'))
dx' }  dx
\end{equation}

This equation has the following asymptotes:
\begin{equation}
G =  \int_{0}^{z} \frac{(z-x)^3 \exp(-G)}
{1+0}  dx
\end{equation}
when $a \rightarrow \infty$ and
\begin{equation}
G =  \int_{0}^{z} \frac{(z-x)^3 \exp(-G)}
{ \frac{1}{c} \int_{0}^{z} \exp(-G(x'))
dx' }  dx
\end{equation}
when $a=0$.

The value of $c=0.08$ is determined in order to have the
approximate coincidence
of the  characteristic
    lengths of the spectrums in these two limit cases.  Our qualitative
analysis shows
that the form of the spectrum
$f \sim \exp(-G)$ will be approximately one and the same for
all values of the parameter $a$. The numerical results shown in the upper
part of Figure 2 confirm this conclusion. We see that the curves
$f_{(a=0)}$ and  $f_{(a=\infty)}$ practically coincide.

Certainly, the integral in the denominator can influence on the process
in another manner. It can attain some value and violates the correct
rescaling of the spectrum to the standard length. The maximum of such
an influence will be attained at $a \sim c$. This case is drawn also and
one can see that in this situation
the flat region of the spectrum has the length of the flat region of the
spectrum in the situation $a=\infty$ and the inclined region has the slope
as the inclined region in the situation $a = 0$ has. This doesn't lead
as
it is seen from this Figure to any essential deviation of the form of
the spectrum.

In the lower part of this Figure the boundaries for the spectrum from below
$f_{bel}$ and from above $f_{ab}$ for $c=0.08$ and arbitrary values of
$a$
are drawn. One can see
that the form of the spectrum is the approximate universal function.

One can investigate now the situation with some finite value of $w_+$. The
analogous
transformations lead to the following equation
\begin{equation}
G  = \int_{0}^{z} \frac{(z-x)^3 \exp(-G)}
{\frac{a}{a+c}+ \frac{1}{a+c} \int_{0}^{z} \exp(-G(x'))
dx' }
[1-\exp(-(a+const \int_0^{x} \exp(-G(x')) dx' ) \exp(w_+)   ]
dx
\end{equation}

For $w_+ = -\infty$ one has the homogeneous situation.
For $w_+ = \infty$ one has the already analyzed situation which as it
has been proved lies not far from the homogeneous one. So, we have the
same limits as the boundaries and can assume that here the spectrum is
also near the universal one.

Note that the physical reason of this universality is the sharp form
of the function $(z-x)^3$ which induces the approximate rectangular form
of the spectrum.

Now
we return to the previous definitions of the parameters made before the
integration
over $w$.
Let us explicitly extract the intermediate and the active regions.
One can see that for the two different activities $w_1$ and $w_2$ the
following
equation is valid
\begin{equation}
\frac{\ln (\theta(w_1,x))}{\exp(w_1)}
=
\frac{\ln (\theta(w_2,x))}{\exp(w_2)}
\end{equation}
The same is valid for the final values. The value of
$\frac{\ln (\theta(w,x))}{\exp(w)}$ is invariant for the  different activities
acting in one and the same process. So, one can put $w=w_{++}$ as the
boundary between these two regions, where $w_{++}$ is equal to $2$ or
$3$.
In the same manner we can separate the region of the unactive centers by
the boundary $w_{--} =  - w_{++}$. One can neglect the substance in the
droplets
on the unactive centers and get\footnote{The absence of the active and
intermediate regions ($w_{+} < w_{--}$) means that condensation occurs
in the pseudo homogeneous way.}
\begin{equation}
G = \int_{w_{--}}^{w_{+}} g(w)  dw
\end{equation}

We denote by $n_{\infty} G_{+}$ the number of the condensing substance
molecules in the droplets formed on the active centers. The system of
the condensation
equations can be rewritten as
$$
g = a_0 \exp(w) \int_{0}^{z} (z-x)^3 \exp(-G) dx
$$
$$
G = \int_{-\infty}^{w_{++}} g(w)  dw  + G_{+}
$$
$$
\theta = \exp( - a_1 \exp(w) \int_{0}^{z} \exp(-G) dx )
$$
Now one can write some expression for $G_{+}$
\begin{equation}
G_{+} = \frac{\eta_{*}}{n_{\infty}} (w_{+} - w_{++})   z^3
\end{equation}
The system of the condensation equations  is the following one
$$
g = a_0 \exp(w) \int_{0}^{z} (z-x)^3 \exp(-G) dx
$$
$$
G = \int_{-\infty}^{w_{++}} g(w)  dw  +
\frac{\eta_{*}}{n_{\infty}} (w_{+} - w_{++})   z^3
$$
$$
\theta = \exp( - a_1 \exp(w) \int_{0}^{z} \exp(-G) dx )
$$
The value of $w_{++}$ is  universal, but the coefficient in the term
$\frac{\eta_{*}}{n_{\infty}} (w_{+} - w_{++})   z^3$ depends on parameters.

Defining
\begin{equation}
G_{-} =\int_{-\infty}^{w_{++}} g(w)  dw
\end{equation}
one can propose the following estimate
\begin{equation}
G_{-} \leq ( w_{++} -  w_{--} ) \frac{\eta_{*}}{n_{\infty}} z^3
\end{equation}
When $$w_{+} - w_{++} \gg w_{++} -  w_{--} $$
one can approximately get
$$
G = \frac{\eta_{*}}{n_{\infty}} (w_{+} - w_{++})    z^3
$$
Then the first equation isn't necessary at all and  one needn't to consider
it. So we needn't to put $a_0$ to $1$ and can use an arbitrary scale to
cancel another constant. The system looks like
$$
\theta = \exp( - a_1 \exp(w) \int_{0}^{z} \exp(-G) dx )
$$
$$
G = \frac{\eta_{*}}{n_{\infty}} (w_{+} - w_{++}) z^3
$$
After the rescaling of $x, z$ one can put $ \frac{\eta_{*}}{n_{\infty}} (w_{+}
-
w_{++})
= 1 $ and the system looks like the universal expression:
$$
\theta = \exp( -  \frac{\ln 2}{\int_{0}^{\infty} \exp(-z^3) dx}
\exp(w) \int_{0}^{z} \exp(-z^3) dx )
$$
Note that this law takes place only in the case of the "wide spectrum".

The behavior of $\theta$ as a function of $z$ for the different values of
$w$ are shown by  Figure 3.

Despite the  accurate character of the methods discussed above these methods
don't allow the clear interpretation. That's why we pay attention to the
simple method presented below.
It can be applied for rather an arbitrary conduction of $\eta_{*}$ as
the function of $w$.

\section{Monodisperse approximation}

The level of the  supersaturation
leading to the cut-off of the spectrum by the exhaustion of
the substance is practically one and the same for all sorts of the droplets
(all sort of the centers).
Let us see the droplets of what sizes play the leading role in this cut-off.
Analyzing the subintegral expression in the equation for  $g$
we realize that this subintegral expression
is the very sharp function of $x$. It is less than the function
\begin{equation}
s_{bel} = \Theta(z-x) (z-x)^3
\end{equation}
and greater than the function
\begin{equation}
s_{ab} = \Theta(z-x) (z-x)^3 \exp(-\frac{\Gamma
\int f_* \eta_{tot}(w) \exp(\lambda w) dw }{\Phi_{*}  }\frac{z^4}{4} )
\exp(-f_* \exp(\lambda w) n_{\infty} z)
\end{equation}

As far as the interruption
of the process of the droplets formation by the fall of the supersaturation
is investigated we shall
describe the pseudohomogeneous situation. Later the generalization
will be presented.
Let us extract the approximation for this function. In other words we must
extract the region of the droplets of the
sizes essential in the vapor consumption.
This consumption in its turn is essential when
\begin{equation}
x
\approx
\Delta x
\end{equation}
where $\Delta x$ is the characteristic size of the cut-off. This value can
be introduced due to the frontal character of the back side of the spectrum.
Certainly, this region must have the sizes rather small in comparison with
$\Delta x$ because all iteration procedures in the homogeneous
decay
are based on the fact
that the droplets formed at the almost ideal supersaturation  determine
 the process
of the spectrum formation.
As for the differential halfwidth $\delta_{1/2}$ we have the following
expression
\begin{equation}
\delta_{1/2}x = (1-\frac{1}{2^{1/3}}) x
\end{equation}
As for the integral halfwidth $\Delta_{1/2}x$, it is obtained from
the following equation
\begin{equation}
N_{ess}x^3 = f_{*}\frac{x^4}{4 } \exp(\lambda w) n_{\infty}
\end{equation}
where $N_{ess} $ is the number of the essential droplets obtained as
$N_{ess}=f_{*}\exp(\lambda w)\Delta_{1/2}x n_{\infty}$ which gives
\begin{equation}
\Delta_{1/2}x = \frac{1}{4} x
\end{equation}
and practically coincides with $\delta_{1/2}x$.
So, the subintegral function $s$ is now split into the essential part where
$x \leq \frac{\Delta x}{4}$ and the tail where $x \geq
 \frac{\Delta x}{4}$.
We shall neglect the tail and use due to rather a small size of
the essential region
the monodisperse approximation for droplets
formed in this region.
As the result we obtain the approximation for $g(x)$
\begin{equation}
g(z) = \frac{N(z/4)}{n_{\infty}}z^3
\end{equation}
where $N(z/4)$ is the number of droplets appeared from $x=0$ till $x=z/4$.
As far as
spectrum is  cut off by the supersaturation exhaustion in the frontal (sharp)
manner
 the value of $g$ is unessential
before $z=\Delta x$
as a small one. After the
moment of the cut-off
it is also unessential  as there is no formation of the droplets. So
instead of the previous approximation we can use
\begin{equation}
g(z) = \frac{N(\Delta x/4)}{n_{\infty}}z^3
\end{equation}
exhaustion of the heterogeneous centers makes the subintegral function
more sharp and the monodisperse approximation  becomes  at
$\Delta x$ even better than in the pseudo homogeneous situation.
 But the exhaustion of the heterogeneous centers makes the
coordinate of the supersaturation cut-off greater than $\Delta x$ and
the monodisperse approximation becomes even more better at the moment
of the supersaturation
cut-off. Certainly, we must use $ N(\Delta x/4)$ calculated with the account
of
the exhaustion of the heterogeneous centers.

The concluding remarks concern the fact that we can  obtain $ N(\Delta x/4)$
by the solution of the equations for the
separate condensation process  because
we need the lowest cut-off length. This length is given without
the cross influence
taking into account due to the frontal character of the back side of
the spectrum.

\section{Equation on the parameters}

We shall return to the initial system and introduce there a new approximation.
The system of the condensation equations will be the following one
\begin{equation} \label{e1}
g=
\exp(w) \int_{0}^{z} (z-x)^3 \exp(-
\frac{\Gamma}{\Phi_*} \frac{N^{tot}(\Delta x/4)}{n_{\infty}}z^3)
\theta dx
\end{equation}
\begin{equation}
G= \int_{-\infty}^{0} g dw
\end{equation}
\begin{equation}\label{e3}
\theta = \exp [-A \exp(w) \int_{0}^{z}
\exp(- \frac{\Gamma}{\Phi_*} \frac{N^{tot}(\Delta x/4)}{n_{\infty}}z^3 ) dx]
\end{equation}
where
\begin{equation}\label{e5}
N^{tot}(\Delta x/4) = \int_{-\infty}^{0} N_{tot}(\Delta x/4) dw
\ \ \ \
N_{tot} (\Delta x / 4) = \eta_{tot} (1 - \theta(\Delta x / 4 ) )
\end{equation}
As far as we use the monodisperse approximation we have derived the
 behavior of the supersaturation and need only the parameters of such
a behavior. So we can leave  integral system (\ref{e1}) -
(\ref{e5}) and reformulate it in the terms
of
$N^{tot}(\Delta x/4)
$, i.e. use only (\ref{e3}) - (\ref{e5}).
As we have seen the value of $\Delta x$ is defined only by
the supersaturation cut-off. Hence, we have
\begin{equation}
\int_{0}^{\frac{\Delta x}{4}} \exp(-
\frac{\Gamma}{\Phi_*}
\frac{N^{tot}(\Delta x/4)}{n_{\infty}}z^3)
dz
= (
\frac{\Phi_*}{\Gamma}
\frac{n_{\infty}}{N^{tot}(\Delta x/4)})^{1/3} E
\end{equation}
where
\begin{equation}
E = \int_{0}^{1/4} \exp(-z^3) dz \approx 0.24
\end{equation}
We obtain the following system
\begin{equation} \label{e8}
\theta (\Delta x /4 ) = \exp[-A \exp(w)
(
\frac{\Phi_*}{\Gamma}
\frac{n_{\infty}}{N^{tot}(\Delta x/4)})^{1/3} E ]
\end{equation}
\begin{equation} \label{e9}
N^{tot}(\Delta x/4) = \int_{-\infty}^{0} dw ( 1 -
\exp[-A \exp(w)
(
\frac{\Phi_*}{\Gamma}
\frac{n_{\infty}}{N^{tot}(\Delta x/4)})^{1/3} E ] )
\eta_*
\end{equation}
The last equation is the ordinary algebraic equation with
the well known methods of the solution. The integral can be taken according to
the
procedure after (\ref{c10}).

When $\eta_{tot}$ is some function  equation (\ref{e9}) remains
the algebraic
one.  The number of the heterogeneous centers is obtained on the base of
$N^{tot}
(\frac{\Delta x}{4})$  and the value of $g$ is obtained due to
(\ref{e1}).

\section{Concluding remarks}

The description of the further periods can be given with the help of
the direct applications of the monodisperse approximation as in the case of
the decay of
the homogeneous
metastable phase. We have the following equation
for the hydrodynamic isolated element
\begin{equation}
\frac{\tau}{\alpha} \frac{dz}{dt} = \zeta = \Phi_{*} -  N^{tot}_r
\frac{z^3}{n_{\infty}}
\end{equation}
where $N^{tot}_r$ is the total number of droplets.
The last equation
 can be easily integrated as far as the  r.h.s. doesn't depend on time. So
there are no problems in the process description until the coalescence.

\pagebreak

\begin{picture}(350,470)
\put(51,350){.}
\put(52,350){.}
\put(53,350){.}
\put(54,350){.}
\put(55,350){.}
\put(56,350){.}
\put(57,350){.}
\put(58,350){.}
\put(59,350){.}
\put(60,350){.}
\put(61,350){.}
\put(62,350){.}
\put(63,350){.}
\put(64,350){.}
\put(65,350){.}
\put(66,350){.}
\put(67,350){.}
\put(68,350){.}
\put(69,350){.}
\put(70,350){.}
\put(71,350){.}
\put(72,350){.}
\put(73,350){.}
\put(74,350){.}
\put(75,350){.}
\put(76,350){.}
\put(77,350){.}
\put(78,350){.}
\put(79,350){.}
\put(80,350){.}
\put(81,349){.}
\put(82,349){.}
\put(83,349){.}
\put(84,349){.}
\put(85,349){.}
\put(86,349){.}
\put(87,349){.}
\put(88,349){.}
\put(89,349){.}
\put(90,348){.}
\put(91,348){.}
\put(92,348){.}
\put(93,348){.}
\put(94,348){.}
\put(95,347){.}
\put(96,347){.}
\put(97,347){.}
\put(98,347){.}
\put(99,346){.}
\put(100,346){.}
\put(101,346){.}
\put(102,345){.}
\put(103,345){.}
\put(104,344){.}
\put(105,344){.}
\put(106,343){.}
\put(107,343){.}
\put(108,342){.}
\put(109,342){.}
\put(110,341){.}
\put(111,341){.}
\put(112,340){.}
\put(113,339){.}
\put(114,339){.}
\put(115,338){.}
\put(116,337){.}
\put(117,337){.}
\put(118,336){.}
\put(119,335){.}
\put(120,334){.}
\put(121,333){.}
\put(122,332){.}
\put(123,331){.}
\put(124,330){.}
\put(125,329){.}
\put(126,328){.}
\put(127,327){.}
\put(128,325){.}
\put(129,324){.}
\put(130,323){.}
\put(131,322){.}
\put(132,320){.}
\put(133,319){.}
\put(134,317){.}
\put(135,316){.}
\put(136,314){.}
\put(137,313){.}
\put(138,311){.}
\put(139,309){.}
\put(140,307){.}
\put(141,306){.}
\put(142,304){.}
\put(143,302){.}
\put(144,300){.}
\put(145,298){.}
\put(146,296){.}
\put(147,294){.}
\put(148,292){.}
\put(149,289){.}
\put(150,287){.}
\put(151,285){.}
\put(152,283){.}
\put(153,280){.}
\put(154,278){.}
\put(155,275){.}
\put(156,273){.}
\put(157,270){.}
\put(158,268){.}
\put(159,265){.}
\put(160,262){.}
\put(161,260){.}
\put(162,257){.}
\put(163,254){.}
\put(164,251){.}
\put(165,249){.}
\put(166,246){.}
\put(167,243){.}
\put(168,240){.}
\put(169,237){.}
\put(170,234){.}
\put(171,231){.}
\put(172,228){.}
\put(173,225){.}
\put(174,222){.}
\put(175,218){.}
\put(176,215){.}
\put(177,212){.}
\put(178,209){.}
\put(179,206){.}
\put(180,203){.}
\put(181,199){.}
\put(182,196){.}
\put(183,193){.}
\put(184,190){.}
\put(185,187){.}
\put(186,184){.}
\put(187,180){.}
\put(188,177){.}
\put(189,174){.}
\put(190,171){.}
\put(191,168){.}
\put(192,165){.}
\put(193,162){.}
\put(194,158){.}
\put(195,155){.}
\put(196,152){.}
\put(197,149){.}
\put(198,146){.}
\put(199,144){.}
\put(200,141){.}
\put(201,138){.}
\put(202,135){.}
\put(203,132){.}
\put(204,129){.}
\put(205,127){.}
\put(206,124){.}
\put(207,121){.}
\put(208,119){.}
\put(209,116){.}
\put(210,114){.}
\put(211,111){.}
\put(212,109){.}
\put(213,107){.}
\put(214,104){.}
\put(215,102){.}
\put(216,100){.}
\put(217,98){.}
\put(218,96){.}
\put(219,94){.}
\put(220,92){.}
\put(221,90){.}
\put(222,88){.}
\put(223,86){.}
\put(224,85){.}
\put(225,83){.}
\put(226,81){.}
\put(227,80){.}
\put(228,78){.}
\put(229,77){.}
\put(230,76){.}
\put(231,74){.}
\put(232,73){.}
\put(233,72){.}
\put(234,70){.}
\put(235,69){.}
\put(236,68){.}
\put(237,67){.}
\put(238,66){.}
\put(239,65){.}
\put(240,64){.}
\put(241,63){.}
\put(242,63){.}
\put(243,62){.}
\put(244,61){.}
\put(245,60){.}
\put(246,60){.}
\put(247,59){.}
\put(248,58){.}
\put(249,58){.}
\put(250,57){.}
\put(251,57){.}
\put(252,56){.}
\put(253,56){.}
\put(254,55){.}
\put(255,55){.}
\put(256,55){.}
\put(257,54){.}
\put(258,54){.}
\put(259,54){.}
\put(260,53){.}
\put(261,53){.}
\put(262,53){.}
\put(263,53){.}
\put(264,52){.}
\put(265,52){.}
\put(266,52){.}
\put(267,52){.}
\put(268,52){.}
\put(269,52){.}
\put(270,51){.}
\put(271,51){.}
\put(272,51){.}
\put(273,51){.}
\put(274,51){.}
\put(275,51){.}
\put(276,51){.}
\put(277,51){.}
\put(10,50){\vector(1,0){350}}
\put(10,50){\vector(0,1){350}}
\put(15,400){$f/f_{max}$}
\put(15,350){$1$}
\put(50,35){0}
\put(150,35){1}
\put(390,35){$z$}
\put(150,10){$Figure \  1$ }
\end{picture}

$$Form \ of \ the \ spectrum \ of \ the \ universal \ solution.$$

\pagebreak

\begin{picture}(350,470)
\put(52,250){.}
\put(52,250){.}
\put(55,250){.}
\put(55,250){.}
\put(58,250){.}
\put(58,250){.}
\put(61,250){.}
\put(61,250){.}
\put(64,250){.}
\put(64,250){.}
\put(67,250){.}
\put(67,250){.}
\put(70,250){.}
\put(70,250){.}
\put(73,250){.}
\put(73,250){.}
\put(76,250){.}
\put(76,250){.}
\put(79,250){.}
\put(79,250){.}
\put(82,250){.}
\put(82,250){.}
\put(85,250){.}
\put(85,249){.}
\put(88,250){.}
\put(88,249){.}
\put(91,250){.}
\put(91,249){.}
\put(94,250){.}
\put(94,249){.}
\put(97,250){.}
\put(97,249){.}
\put(100,249){.}
\put(100,248){.}
\put(103,249){.}
\put(103,248){.}
\put(106,249){.}
\put(106,247){.}
\put(109,249){.}
\put(109,247){.}
\put(112,249){.}
\put(112,246){.}
\put(115,248){.}
\put(115,246){.}
\put(118,248){.}
\put(118,245){.}
\put(121,248){.}
\put(121,244){.}
\put(124,247){.}
\put(124,244){.}
\put(127,247){.}
\put(127,243){.}
\put(130,246){.}
\put(130,242){.}
\put(133,246){.}
\put(133,241){.}
\put(136,245){.}
\put(136,240){.}
\put(139,244){.}
\put(139,239){.}
\put(142,244){.}
\put(142,237){.}
\put(145,243){.}
\put(145,236){.}
\put(148,242){.}
\put(148,234){.}
\put(151,241){.}
\put(151,233){.}
\put(154,240){.}
\put(154,231){.}
\put(157,239){.}
\put(157,230){.}
\put(160,238){.}
\put(160,228){.}
\put(163,236){.}
\put(163,226){.}
\put(166,235){.}
\put(166,224){.}
\put(169,234){.}
\put(169,222){.}
\put(172,232){.}
\put(172,220){.}
\put(175,231){.}
\put(175,217){.}
\put(178,229){.}
\put(178,215){.}
\put(181,227){.}
\put(181,212){.}
\put(184,226){.}
\put(184,210){.}
\put(187,224){.}
\put(187,207){.}
\put(190,222){.}
\put(190,204){.}
\put(193,220){.}
\put(193,202){.}
\put(196,218){.}
\put(196,199){.}
\put(199,215){.}
\put(199,196){.}
\put(202,213){.}
\put(202,193){.}
\put(205,211){.}
\put(205,190){.}
\put(208,208){.}
\put(208,186){.}
\put(211,206){.}
\put(211,183){.}
\put(214,203){.}
\put(214,180){.}
\put(217,201){.}
\put(217,177){.}
\put(220,198){.}
\put(220,173){.}
\put(223,195){.}
\put(223,170){.}
\put(226,192){.}
\put(226,166){.}
\put(229,189){.}
\put(229,163){.}
\put(232,186){.}
\put(232,159){.}
\put(235,183){.}
\put(235,156){.}
\put(238,180){.}
\put(238,152){.}
\put(241,177){.}
\put(241,149){.}
\put(244,174){.}
\put(244,145){.}
\put(247,171){.}
\put(247,142){.}
\put(250,168){.}
\put(250,138){.}
\put(253,164){.}
\put(253,135){.}
\put(256,161){.}
\put(256,131){.}
\put(259,158){.}
\put(259,128){.}
\put(262,154){.}
\put(262,124){.}
\put(265,151){.}
\put(265,121){.}
\put(268,148){.}
\put(268,117){.}
\put(271,144){.}
\put(271,113){.}
\put(274,141){.}
\put(274,109){.}
\put(277,138){.}
\put(277,105){.}
\put(280,135){.}
\put(280,102){.}
\put(283,131){.}
\put(283,98){.}
\put(286,128){.}
\put(286,95){.}
\put(289,125){.}
\put(289,91){.}
\put(292,122){.}
\put(292,88){.}
\put(295,119){.}
\put(295,85){.}
\put(298,116){.}
\put(298,82){.}
\put(301,113){.}
\put(301,80){.}
\put(304,110){.}
\put(304,77){.}
\put(307,107){.}
\put(307,75){.}
\put(310,104){.}
\put(310,72){.}
\put(313,101){.}
\put(313,70){.}
\put(316,99){.}
\put(316,68){.}
\put(319,96){.}
\put(319,66){.}
\put(322,93){.}
\put(322,65){.}
\put(325,91){.}
\put(325,63){.}
\put(328,89){.}
\put(328,62){.}
\put(331,86){.}
\put(331,60){.}
\put(334,84){.}
\put(334,59){.}
\put(337,82){.}
\put(337,58){.}
\put(340,80){.}
\put(340,57){.}
\put(343,78){.}
\put(343,56){.}
\put(346,76){.}
\put(346,55){.}
\put(349,74){.}
\put(349,55){.}

\put(350,75){$f_{ab}$}

\put(350,60){$f_{bel}$}

\put(52,500){.}
\put(55,500){.}
\put(58,500){.}
\put(61,500){.}
\put(64,500){.}
\put(67,500){.}
\put(70,500){.}
\put(73,500){.}
\put(76,500){.}
\put(79,500){.}
\put(82,500){.}
\put(85,499){.}
\put(88,499){.}
\put(91,499){.}
\put(94,499){.}
\put(97,499){.}
\put(100,498){.}
\put(103,498){.}
\put(106,497){.}
\put(109,497){.}
\put(112,496){.}
\put(115,496){.}
\put(118,495){.}
\put(121,494){.}
\put(124,494){.}
\put(127,493){.}
\put(130,492){.}
\put(133,491){.}
\put(136,490){.}
\put(139,489){.}
\put(142,487){.}
\put(145,486){.}
\put(148,484){.}
\put(151,483){.}
\put(154,481){.}
\put(157,480){.}
\put(160,478){.}
\put(163,476){.}
\put(166,474){.}
\put(169,472){.}
\put(172,470){.}
\put(175,467){.}
\put(178,465){.}
\put(181,462){.}
\put(184,460){.}
\put(187,457){.}
\put(190,454){.}
\put(193,452){.}
\put(196,449){.}
\put(199,446){.}
\put(202,443){.}
\put(205,440){.}
\put(208,436){.}
\put(211,433){.}
\put(214,430){.}
\put(217,427){.}
\put(220,423){.}
\put(223,420){.}
\put(226,416){.}
\put(229,413){.}
\put(232,409){.}
\put(235,406){.}
\put(238,402){.}
\put(241,399){.}
\put(244,395){.}
\put(247,392){.}
\put(250,388){.}
\put(253,385){.}
\put(256,381){.}
\put(259,378){.}
\put(262,374){.}
\put(265,371){.}
\put(268,368){.}
\put(271,365){.}
\put(274,361){.}
\put(277,358){.}
\put(280,355){.}
\put(283,352){.}
\put(286,350){.}
\put(289,347){.}
\put(292,344){.}
\put(295,341){.}
\put(298,339){.}
\put(301,337){.}
\put(304,334){.}
\put(307,332){.}
\put(310,330){.}
\put(313,328){.}
\put(316,326){.}
\put(319,324){.}
\put(322,322){.}
\put(325,321){.}
\put(328,319){.}
\put(331,317){.}
\put(334,316){.}
\put(337,315){.}
\put(340,313){.}
\put(343,312){.}
\put(346,311){.}
\put(349,310){.}

\put(350,310){$f_{(a \rightarrow 0)}$}

\put(52,500){.}
\put(55,500){.}
\put(58,500){.}
\put(61,500){.}
\put(64,500){.}
\put(67,500){.}
\put(70,500){.}
\put(73,500){.}
\put(76,500){.}
\put(79,500){.}
\put(82,500){.}
\put(85,500){.}
\put(88,500){.}
\put(91,500){.}
\put(94,500){.}
\put(97,500){.}
\put(100,499){.}
\put(103,499){.}
\put(106,499){.}
\put(109,499){.}
\put(112,499){.}
\put(115,498){.}
\put(118,498){.}
\put(121,498){.}
\put(124,497){.}
\put(127,497){.}
\put(130,496){.}
\put(133,495){.}
\put(136,495){.}
\put(139,494){.}
\put(142,493){.}
\put(145,492){.}
\put(148,491){.}
\put(151,490){.}
\put(154,489){.}
\put(157,488){.}
\put(160,486){.}
\put(163,485){.}
\put(166,483){.}
\put(169,481){.}
\put(172,479){.}
\put(175,477){.}
\put(178,475){.}
\put(181,473){.}
\put(184,471){.}
\put(187,468){.}
\put(190,466){.}
\put(193,463){.}
\put(196,460){.}
\put(199,457){.}
\put(202,454){.}
\put(205,451){.}
\put(208,447){.}
\put(211,444){.}
\put(214,440){.}
\put(217,437){.}
\put(220,433){.}
\put(223,429){.}
\put(226,425){.}
\put(229,421){.}
\put(232,417){.}
\put(235,413){.}
\put(238,409){.}
\put(241,404){.}
\put(244,400){.}
\put(247,396){.}
\put(250,392){.}
\put(253,387){.}
\put(256,383){.}
\put(259,379){.}
\put(262,375){.}
\put(265,371){.}
\put(268,367){.}
\put(271,363){.}
\put(274,359){.}
\put(277,355){.}
\put(280,352){.}
\put(283,348){.}
\put(286,345){.}
\put(289,341){.}
\put(292,338){.}
\put(295,335){.}
\put(298,332){.}
\put(301,330){.}
\put(304,327){.}
\put(307,325){.}
\put(310,322){.}
\put(313,320){.}
\put(316,318){.}
\put(319,316){.}
\put(322,315){.}
\put(325,313){.}
\put(328,312){.}
\put(331,310){.}
\put(334,309){.}
\put(337,308){.}
\put(340,307){.}
\put(343,306){.}
\put(346,305){.}
\put(349,305){.}

\put(300,310){$f_{(a \rightarrow \infty)}$}

\put(52,500){.}
\put(55,500){.}
\put(58,500){.}
\put(61,500){.}
\put(64,500){.}
\put(67,500){.}
\put(70,500){.}
\put(73,500){.}
\put(76,500){.}
\put(79,500){.}
\put(82,500){.}
\put(85,500){.}
\put(88,500){.}
\put(91,500){.}
\put(94,500){.}
\put(97,499){.}
\put(100,499){.}
\put(103,499){.}
\put(106,499){.}
\put(109,499){.}
\put(112,498){.}
\put(115,498){.}
\put(118,498){.}
\put(121,497){.}
\put(124,497){.}
\put(127,496){.}
\put(130,496){.}
\put(133,495){.}
\put(136,495){.}
\put(139,494){.}
\put(142,493){.}
\put(145,493){.}
\put(148,492){.}
\put(151,491){.}
\put(154,490){.}
\put(157,489){.}
\put(160,487){.}
\put(163,486){.}
\put(166,485){.}
\put(169,484){.}
\put(172,482){.}
\put(175,481){.}
\put(178,479){.}
\put(181,477){.}
\put(184,476){.}
\put(187,474){.}
\put(190,472){.}
\put(193,470){.}
\put(196,468){.}
\put(199,465){.}
\put(202,463){.}
\put(205,461){.}
\put(208,458){.}
\put(211,456){.}
\put(214,453){.}
\put(217,451){.}
\put(220,448){.}
\put(223,445){.}
\put(226,442){.}
\put(229,439){.}
\put(232,436){.}
\put(235,433){.}
\put(238,430){.}
\put(241,427){.}
\put(244,424){.}
\put(247,421){.}
\put(250,417){.}
\put(253,414){.}
\put(256,411){.}
\put(259,407){.}
\put(262,404){.}
\put(265,401){.}
\put(268,397){.}
\put(271,394){.}
\put(274,391){.}
\put(277,387){.}
\put(280,384){.}
\put(283,381){.}
\put(286,378){.}
\put(289,374){.}
\put(292,371){.}
\put(295,368){.}
\put(298,365){.}
\put(301,362){.}
\put(304,359){.}
\put(307,356){.}
\put(310,353){.}
\put(313,351){.}
\put(316,348){.}
\put(319,345){.}
\put(322,343){.}
\put(325,340){.}
\put(328,338){.}
\put(331,336){.}
\put(334,333){.}
\put(337,331){.}
\put(340,329){.}
\put(343,327){.}
\put(346,325){.}
\put(349,324){.}

\put(350,330){$f_{(a \sim d = 0.08 )}$}

\put(50,50){\vector(1,0){400}}
\put(50,300){\vector(1,0){400}}
\put(50,50){\vector(0,1){220}}
\put(50,300){\vector(0,1){220}}
\put(50,30){0}
\put(50,280){0}
\put(200,30){1}
\put(200,280){1}
\put(360,30){$z$}
\put(360,280){$z$}
\put(150,5){$Figure \  2$}
\end{picture}

$$ Approximate \ universiality \ of \ the \ spectrum.$$

\pagebreak

\begin{picture}(350,470)
\put(51,350){.}
\put(52,349){.}
\put(53,349){.}
\put(54,349){.}
\put(55,348){.}
\put(56,348){.}
\put(57,348){.}
\put(58,347){.}
\put(59,347){.}
\put(60,347){.}
\put(61,347){.}
\put(62,346){.}
\put(63,346){.}
\put(64,346){.}
\put(65,345){.}
\put(66,345){.}
\put(67,345){.}
\put(68,344){.}
\put(69,344){.}
\put(70,344){.}
\put(71,343){.}
\put(72,343){.}
\put(73,343){.}
\put(74,343){.}
\put(75,342){.}
\put(76,342){.}
\put(77,342){.}
\put(78,341){.}
\put(79,341){.}
\put(80,341){.}
\put(81,340){.}
\put(82,340){.}
\put(83,340){.}
\put(84,340){.}
\put(85,339){.}
\put(86,339){.}
\put(87,339){.}
\put(88,338){.}
\put(89,338){.}
\put(90,338){.}
\put(91,338){.}
\put(92,337){.}
\put(93,337){.}
\put(94,337){.}
\put(95,336){.}
\put(96,336){.}
\put(97,336){.}
\put(98,336){.}
\put(99,335){.}
\put(100,335){.}
\put(101,335){.}
\put(102,335){.}
\put(103,334){.}
\put(104,334){.}
\put(105,334){.}
\put(106,333){.}
\put(107,333){.}
\put(108,333){.}
\put(109,333){.}
\put(110,333){.}
\put(111,332){.}
\put(112,332){.}
\put(113,332){.}
\put(114,332){.}
\put(115,331){.}
\put(116,331){.}
\put(117,331){.}
\put(118,331){.}
\put(119,330){.}
\put(120,330){.}
\put(121,330){.}
\put(122,330){.}
\put(123,330){.}
\put(124,329){.}
\put(125,329){.}
\put(126,329){.}
\put(127,329){.}
\put(128,329){.}
\put(129,329){.}
\put(130,328){.}
\put(131,328){.}
\put(132,328){.}
\put(133,328){.}
\put(134,328){.}
\put(135,328){.}
\put(136,327){.}
\put(137,327){.}
\put(138,327){.}
\put(139,327){.}
\put(140,327){.}
\put(141,327){.}
\put(142,327){.}
\put(143,326){.}
\put(144,326){.}
\put(145,326){.}
\put(146,326){.}
\put(147,326){.}
\put(148,326){.}
\put(149,326){.}
\put(150,326){.}
\put(151,325){.}
\put(152,325){.}
\put(153,325){.}
\put(154,325){.}
\put(155,325){.}
\put(156,325){.}
\put(157,325){.}
\put(158,325){.}
\put(159,325){.}
\put(160,325){.}
\put(161,325){.}
\put(162,325){.}
\put(163,324){.}
\put(164,324){.}
\put(165,324){.}
\put(166,324){.}
\put(167,324){.}
\put(168,324){.}
\put(169,324){.}
\put(170,324){.}
\put(171,324){.}
\put(172,324){.}
\put(173,324){.}
\put(174,324){.}
\put(175,324){.}
\put(176,324){.}
\put(177,324){.}
\put(178,324){.}
\put(179,324){.}
\put(180,324){.}
\put(181,324){.}
\put(182,324){.}
\put(183,324){.}
\put(184,324){.}
\put(185,323){.}
\put(186,323){.}
\put(187,323){.}
\put(188,323){.}
\put(189,323){.}
\put(190,323){.}
\put(191,323){.}
\put(192,323){.}
\put(193,323){.}
\put(194,323){.}
\put(195,323){.}
\put(196,323){.}
\put(197,323){.}
\put(198,323){.}
\put(199,323){.}
\put(200,323){.}
\put(201,323){.}
\put(202,323){.}
\put(203,323){.}
\put(204,323){.}
\put(205,323){.}
\put(206,323){.}
\put(207,323){.}
\put(208,323){.}
\put(209,323){.}
\put(210,323){.}
\put(211,323){.}
\put(212,323){.}
\put(213,323){.}
\put(214,323){.}
\put(215,323){.}
\put(216,323){.}
\put(217,323){.}
\put(218,323){.}
\put(219,323){.}
\put(220,323){.}
\put(221,323){.}
\put(222,323){.}
\put(223,323){.}
\put(224,323){.}
\put(225,323){.}
\put(226,323){.}
\put(227,323){.}
\put(228,323){.}
\put(229,323){.}
\put(230,323){.}
\put(231,323){.}
\put(232,323){.}
\put(233,323){.}
\put(234,323){.}
\put(235,323){.}
\put(236,323){.}
\put(237,323){.}
\put(238,323){.}
\put(239,323){.}
\put(240,323){.}
\put(241,323){.}
\put(242,323){.}
\put(243,323){.}
\put(244,323){.}
\put(245,323){.}
\put(246,323){.}
\put(247,323){.}
\put(248,323){.}
\put(249,323){.}
\put(250,323){.}
\put(251,323){.}
\put(252,323){.}
\put(253,323){.}
\put(254,323){.}
\put(255,323){.}
\put(256,323){.}
\put(257,323){.}
\put(258,323){.}
\put(259,323){.}
\put(260,323){.}
\put(261,323){.}
\put(262,323){.}
\put(263,323){.}
\put(264,323){.}
\put(265,323){.}
\put(266,323){.}
\put(267,323){.}
\put(268,323){.}
\put(269,323){.}
\put(270,323){.}
\put(271,323){.}
\put(272,323){.}
\put(273,323){.}
\put(274,323){.}
\put(275,323){.}
\put(276,323){.}
\put(277,323){.}
\put(278,323){.}
\put(279,323){.}
\put(280,323){.}
\put(281,323){.}
\put(282,323){.}
\put(283,323){.}
\put(284,323){.}
\put(285,323){.}
\put(286,323){.}
\put(287,323){.}
\put(288,323){.}
\put(289,323){.}
\put(290,323){.}
\put(291,323){.}
\put(292,323){.}
\put(293,323){.}
\put(294,323){.}
\put(295,323){.}
\put(296,323){.}
\put(297,323){.}
\put(298,323){.}
\put(299,323){.}
\put(300,323){.}
\put(301,323){.}
\put(302,323){.}
\put(303,323){.}
\put(304,323){.}
\put(305,323){.}
\put(306,323){.}
\put(307,323){.}
\put(308,323){.}
\put(309,323){.}
\put(310,323){.}
\put(311,323){.}
\put(312,323){.}
\put(313,323){.}
\put(314,323){.}
\put(315,323){.}
\put(316,323){.}
\put(317,323){.}
\put(318,323){.}
\put(319,323){.}
\put(320,323){.}
\put(321,323){.}
\put(322,323){.}
\put(323,323){.}
\put(324,323){.}
\put(325,323){.}
\put(326,323){.}
\put(327,323){.}
\put(328,323){.}
\put(329,323){.}
\put(330,323){.}
\put(331,323){.}
\put(332,323){.}
\put(333,323){.}
\put(334,323){.}
\put(335,323){.}
\put(336,323){.}
\put(337,323){.}
\put(338,323){.}
\put(339,323){.}
\put(340,323){.}
\put(341,323){.}
\put(342,323){.}
\put(343,323){.}
\put(344,323){.}
\put(345,323){.}
\put(346,323){.}
\put(347,323){.}
\put(348,323){.}
\put(349,323){.}
\put(350,323){.}
\put(351,323){.}
\put(355,330){$w=-2$}
\put(51,349){.}
\put(52,348){.}
\put(53,347){.}
\put(54,347){.}
\put(55,346){.}
\put(56,345){.}
\put(57,344){.}
\put(58,343){.}
\put(59,342){.}
\put(60,342){.}
\put(61,341){.}
\put(62,340){.}
\put(63,339){.}
\put(64,338){.}
\put(65,337){.}
\put(66,337){.}
\put(67,336){.}
\put(68,335){.}
\put(69,334){.}
\put(70,333){.}
\put(71,332){.}
\put(72,332){.}
\put(73,331){.}
\put(74,330){.}
\put(75,329){.}
\put(76,329){.}
\put(77,328){.}
\put(78,327){.}
\put(79,326){.}
\put(80,325){.}
\put(81,325){.}
\put(82,324){.}
\put(83,323){.}
\put(84,322){.}
\put(85,322){.}
\put(86,321){.}
\put(87,320){.}
\put(88,319){.}
\put(89,319){.}
\put(90,318){.}
\put(91,317){.}
\put(92,317){.}
\put(93,316){.}
\put(94,315){.}
\put(95,314){.}
\put(96,314){.}
\put(97,313){.}
\put(98,312){.}
\put(99,312){.}
\put(100,311){.}
\put(101,310){.}
\put(102,310){.}
\put(103,309){.}
\put(104,308){.}
\put(105,308){.}
\put(106,307){.}
\put(107,307){.}
\put(108,306){.}
\put(109,305){.}
\put(110,305){.}
\put(111,304){.}
\put(112,304){.}
\put(113,303){.}
\put(114,303){.}
\put(115,302){.}
\put(116,301){.}
\put(117,301){.}
\put(118,300){.}
\put(119,300){.}
\put(120,299){.}
\put(121,299){.}
\put(122,298){.}
\put(123,298){.}
\put(124,297){.}
\put(125,297){.}
\put(126,296){.}
\put(127,296){.}
\put(128,296){.}
\put(129,295){.}
\put(130,295){.}
\put(131,294){.}
\put(132,294){.}
\put(133,294){.}
\put(134,293){.}
\put(135,293){.}
\put(136,292){.}
\put(137,292){.}
\put(138,292){.}
\put(139,291){.}
\put(140,291){.}
\put(141,291){.}
\put(142,290){.}
\put(143,290){.}
\put(144,290){.}
\put(145,289){.}
\put(146,289){.}
\put(147,289){.}
\put(148,289){.}
\put(149,288){.}
\put(150,288){.}
\put(151,288){.}
\put(152,288){.}
\put(153,287){.}
\put(154,287){.}
\put(155,287){.}
\put(156,287){.}
\put(157,287){.}
\put(158,286){.}
\put(159,286){.}
\put(160,286){.}
\put(161,286){.}
\put(162,286){.}
\put(163,286){.}
\put(164,285){.}
\put(165,285){.}
\put(166,285){.}
\put(167,285){.}
\put(168,285){.}
\put(169,285){.}
\put(170,285){.}
\put(171,284){.}
\put(172,284){.}
\put(173,284){.}
\put(174,284){.}
\put(175,284){.}
\put(176,284){.}
\put(177,284){.}
\put(178,284){.}
\put(179,284){.}
\put(180,284){.}
\put(181,284){.}
\put(182,283){.}
\put(183,283){.}
\put(184,283){.}
\put(185,283){.}
\put(186,283){.}
\put(187,283){.}
\put(188,283){.}
\put(189,283){.}
\put(190,283){.}
\put(191,283){.}
\put(192,283){.}
\put(193,283){.}
\put(194,283){.}
\put(195,283){.}
\put(196,283){.}
\put(197,283){.}
\put(198,283){.}
\put(199,283){.}
\put(200,283){.}
\put(201,283){.}
\put(202,283){.}
\put(203,283){.}
\put(204,283){.}
\put(205,283){.}
\put(206,283){.}
\put(207,283){.}
\put(208,283){.}
\put(209,283){.}
\put(210,283){.}
\put(211,283){.}
\put(212,283){.}
\put(213,283){.}
\put(214,283){.}
\put(215,283){.}
\put(216,283){.}
\put(217,283){.}
\put(218,283){.}
\put(219,283){.}
\put(220,283){.}
\put(221,283){.}
\put(222,283){.}
\put(223,283){.}
\put(224,283){.}
\put(225,283){.}
\put(226,283){.}
\put(227,282){.}
\put(228,282){.}
\put(229,282){.}
\put(230,282){.}
\put(231,282){.}
\put(232,282){.}
\put(233,282){.}
\put(234,282){.}
\put(235,282){.}
\put(236,282){.}
\put(237,282){.}
\put(238,282){.}
\put(239,282){.}
\put(240,282){.}
\put(241,282){.}
\put(242,282){.}
\put(243,282){.}
\put(244,282){.}
\put(245,282){.}
\put(246,282){.}
\put(247,282){.}
\put(248,282){.}
\put(249,282){.}
\put(250,282){.}
\put(251,282){.}
\put(252,282){.}
\put(253,282){.}
\put(254,282){.}
\put(255,282){.}
\put(256,282){.}
\put(257,282){.}
\put(258,282){.}
\put(259,282){.}
\put(260,282){.}
\put(261,282){.}
\put(262,282){.}
\put(263,282){.}
\put(264,282){.}
\put(265,282){.}
\put(266,282){.}
\put(267,282){.}
\put(268,282){.}
\put(269,282){.}
\put(270,282){.}
\put(271,282){.}
\put(272,282){.}
\put(273,282){.}
\put(274,282){.}
\put(275,282){.}
\put(276,282){.}
\put(277,282){.}
\put(278,282){.}
\put(279,282){.}
\put(280,282){.}
\put(281,282){.}
\put(282,282){.}
\put(283,282){.}
\put(284,282){.}
\put(285,282){.}
\put(286,282){.}
\put(287,282){.}
\put(288,282){.}
\put(289,282){.}
\put(290,282){.}
\put(291,282){.}
\put(292,282){.}
\put(293,282){.}
\put(294,282){.}
\put(295,282){.}
\put(296,282){.}
\put(297,282){.}
\put(298,282){.}
\put(299,282){.}
\put(300,282){.}
\put(301,282){.}
\put(302,282){.}
\put(303,282){.}
\put(304,282){.}
\put(305,282){.}
\put(306,282){.}
\put(307,282){.}
\put(308,282){.}
\put(309,282){.}
\put(310,282){.}
\put(311,282){.}
\put(312,282){.}
\put(313,282){.}
\put(314,282){.}
\put(315,282){.}
\put(316,282){.}
\put(317,282){.}
\put(318,282){.}
\put(319,282){.}
\put(320,282){.}
\put(321,282){.}
\put(322,282){.}
\put(323,282){.}
\put(324,282){.}
\put(325,282){.}
\put(326,282){.}
\put(327,282){.}
\put(328,282){.}
\put(329,282){.}
\put(330,282){.}
\put(331,282){.}
\put(332,282){.}
\put(333,282){.}
\put(334,282){.}
\put(335,282){.}
\put(336,282){.}
\put(337,282){.}
\put(338,282){.}
\put(339,282){.}
\put(340,282){.}
\put(341,282){.}
\put(342,282){.}
\put(343,282){.}
\put(344,282){.}
\put(345,282){.}
\put(346,282){.}
\put(347,282){.}
\put(348,282){.}
\put(349,282){.}
\put(350,282){.}
\put(351,282){.}
\put(355,282){$w=-1$}
\put(51,348){.}
\put(52,345){.}
\put(53,343){.}
\put(54,341){.}
\put(55,339){.}
\put(56,336){.}
\put(57,334){.}
\put(58,332){.}
\put(59,330){.}
\put(60,327){.}
\put(61,325){.}
\put(62,323){.}
\put(63,321){.}
\put(64,319){.}
\put(65,317){.}
\put(66,315){.}
\put(67,313){.}
\put(68,311){.}
\put(69,309){.}
\put(70,307){.}
\put(71,305){.}
\put(72,303){.}
\put(73,301){.}
\put(74,299){.}
\put(75,297){.}
\put(76,295){.}
\put(77,293){.}
\put(78,291){.}
\put(79,290){.}
\put(80,288){.}
\put(81,286){.}
\put(82,284){.}
\put(83,282){.}
\put(84,281){.}
\put(85,279){.}
\put(86,277){.}
\put(87,276){.}
\put(88,274){.}
\put(89,272){.}
\put(90,271){.}
\put(91,269){.}
\put(92,267){.}
\put(93,266){.}
\put(94,264){.}
\put(95,263){.}
\put(96,261){.}
\put(97,260){.}
\put(98,258){.}
\put(99,257){.}
\put(100,256){.}
\put(101,254){.}
\put(102,253){.}
\put(103,251){.}
\put(104,250){.}
\put(105,249){.}
\put(106,247){.}
\put(107,246){.}
\put(108,245){.}
\put(109,244){.}
\put(110,242){.}
\put(111,241){.}
\put(112,240){.}
\put(113,239){.}
\put(114,238){.}
\put(115,237){.}
\put(116,236){.}
\put(117,235){.}
\put(118,234){.}
\put(119,232){.}
\put(120,231){.}
\put(121,230){.}
\put(122,230){.}
\put(123,229){.}
\put(124,228){.}
\put(125,227){.}
\put(126,226){.}
\put(127,225){.}
\put(128,224){.}
\put(129,223){.}
\put(130,222){.}
\put(131,222){.}
\put(132,221){.}
\put(133,220){.}
\put(134,219){.}
\put(135,219){.}
\put(136,218){.}
\put(137,217){.}
\put(138,217){.}
\put(139,216){.}
\put(140,215){.}
\put(141,215){.}
\put(142,214){.}
\put(143,214){.}
\put(144,213){.}
\put(145,213){.}
\put(146,212){.}
\put(147,212){.}
\put(148,211){.}
\put(149,211){.}
\put(150,210){.}
\put(151,210){.}
\put(152,209){.}
\put(153,209){.}
\put(154,208){.}
\put(155,208){.}
\put(156,208){.}
\put(157,207){.}
\put(158,207){.}
\put(159,207){.}
\put(160,206){.}
\put(161,206){.}
\put(162,206){.}
\put(163,205){.}
\put(164,205){.}
\put(165,205){.}
\put(166,205){.}
\put(167,204){.}
\put(168,204){.}
\put(169,204){.}
\put(170,204){.}
\put(171,203){.}
\put(172,203){.}
\put(173,203){.}
\put(174,203){.}
\put(175,203){.}
\put(176,203){.}
\put(177,202){.}
\put(178,202){.}
\put(179,202){.}
\put(180,202){.}
\put(181,202){.}
\put(182,202){.}
\put(183,202){.}
\put(184,202){.}
\put(185,201){.}
\put(186,201){.}
\put(187,201){.}
\put(188,201){.}
\put(189,201){.}
\put(190,201){.}
\put(191,201){.}
\put(192,201){.}
\put(193,201){.}
\put(194,201){.}
\put(195,201){.}
\put(196,201){.}
\put(197,201){.}
\put(198,201){.}
\put(199,201){.}
\put(200,200){.}
\put(201,200){.}
\put(202,200){.}
\put(203,200){.}
\put(204,200){.}
\put(205,200){.}
\put(206,200){.}
\put(207,200){.}
\put(208,200){.}
\put(209,200){.}
\put(210,200){.}
\put(211,200){.}
\put(212,200){.}
\put(213,200){.}
\put(214,200){.}
\put(215,200){.}
\put(216,200){.}
\put(217,200){.}
\put(218,200){.}
\put(219,200){.}
\put(220,200){.}
\put(221,200){.}
\put(222,200){.}
\put(223,200){.}
\put(224,200){.}
\put(225,200){.}
\put(226,200){.}
\put(227,200){.}
\put(228,200){.}
\put(229,200){.}
\put(230,200){.}
\put(231,200){.}
\put(232,200){.}
\put(233,200){.}
\put(234,200){.}
\put(235,200){.}
\put(236,200){.}
\put(237,200){.}
\put(238,200){.}
\put(239,200){.}
\put(240,200){.}
\put(241,200){.}
\put(242,200){.}
\put(243,200){.}
\put(244,200){.}
\put(245,200){.}
\put(246,200){.}
\put(247,200){.}
\put(248,200){.}
\put(249,200){.}
\put(250,200){.}
\put(251,200){.}
\put(252,200){.}
\put(253,200){.}
\put(254,200){.}
\put(255,200){.}
\put(256,200){.}
\put(257,200){.}
\put(258,200){.}
\put(259,200){.}
\put(260,200){.}
\put(261,200){.}
\put(262,200){.}
\put(263,200){.}
\put(264,200){.}
\put(265,200){.}
\put(266,200){.}
\put(267,200){.}
\put(268,200){.}
\put(269,200){.}
\put(270,200){.}
\put(271,200){.}
\put(272,200){.}
\put(273,200){.}
\put(274,200){.}
\put(275,200){.}
\put(276,200){.}
\put(277,200){.}
\put(278,200){.}
\put(279,200){.}
\put(280,200){.}
\put(281,200){.}
\put(282,200){.}
\put(283,200){.}
\put(284,200){.}
\put(285,200){.}
\put(286,200){.}
\put(287,200){.}
\put(288,200){.}
\put(289,200){.}
\put(290,200){.}
\put(291,200){.}
\put(292,200){.}
\put(293,200){.}
\put(294,200){.}
\put(295,200){.}
\put(296,200){.}
\put(297,200){.}
\put(298,200){.}
\put(299,200){.}
\put(300,200){.}
\put(301,200){.}
\put(302,200){.}
\put(303,200){.}
\put(304,200){.}
\put(305,200){.}
\put(306,200){.}
\put(307,200){.}
\put(308,200){.}
\put(309,200){.}
\put(310,200){.}
\put(311,200){.}
\put(312,200){.}
\put(313,200){.}
\put(314,200){.}
\put(315,200){.}
\put(316,200){.}
\put(317,200){.}
\put(318,200){.}
\put(319,200){.}
\put(320,200){.}
\put(321,200){.}
\put(322,200){.}
\put(323,200){.}
\put(324,200){.}
\put(325,200){.}
\put(326,200){.}
\put(327,200){.}
\put(328,200){.}
\put(329,200){.}
\put(330,200){.}
\put(331,200){.}
\put(332,200){.}
\put(333,200){.}
\put(334,200){.}
\put(335,200){.}
\put(336,200){.}
\put(337,200){.}
\put(338,200){.}
\put(339,200){.}
\put(340,200){.}
\put(341,200){.}
\put(342,200){.}
\put(343,200){.}
\put(344,200){.}
\put(345,200){.}
\put(346,200){.}
\put(347,200){.}
\put(348,200){.}
\put(349,200){.}
\put(350,200){.}
\put(351,200){.}
\put(355,200){$w=0$}
\put(51,344){.}
\put(52,338){.}
\put(53,331){.}
\put(54,326){.}
\put(55,320){.}
\put(56,314){.}
\put(57,309){.}
\put(58,303){.}
\put(59,298){.}
\put(60,293){.}
\put(61,288){.}
\put(62,283){.}
\put(63,278){.}
\put(64,273){.}
\put(65,268){.}
\put(66,264){.}
\put(67,259){.}
\put(68,255){.}
\put(69,251){.}
\put(70,246){.}
\put(71,242){.}
\put(72,238){.}
\put(73,234){.}
\put(74,231){.}
\put(75,227){.}
\put(76,223){.}
\put(77,220){.}
\put(78,216){.}
\put(79,213){.}
\put(80,209){.}
\put(81,206){.}
\put(82,203){.}
\put(83,200){.}
\put(84,197){.}
\put(85,194){.}
\put(86,191){.}
\put(87,188){.}
\put(88,185){.}
\put(89,183){.}
\put(90,180){.}
\put(91,178){.}
\put(92,175){.}
\put(93,173){.}
\put(94,170){.}
\put(95,168){.}
\put(96,166){.}
\put(97,164){.}
\put(98,161){.}
\put(99,159){.}
\put(100,157){.}
\put(101,155){.}
\put(102,153){.}
\put(103,152){.}
\put(104,150){.}
\put(105,148){.}
\put(106,146){.}
\put(107,145){.}
\put(108,143){.}
\put(109,141){.}
\put(110,140){.}
\put(111,138){.}
\put(112,137){.}
\put(113,135){.}
\put(114,134){.}
\put(115,133){.}
\put(116,131){.}
\put(117,130){.}
\put(118,129){.}
\put(119,128){.}
\put(120,127){.}
\put(121,125){.}
\put(122,124){.}
\put(123,123){.}
\put(124,122){.}
\put(125,121){.}
\put(126,120){.}
\put(127,119){.}
\put(128,118){.}
\put(129,118){.}
\put(130,117){.}
\put(131,116){.}
\put(132,115){.}
\put(133,114){.}
\put(134,114){.}
\put(135,113){.}
\put(136,112){.}
\put(137,111){.}
\put(138,111){.}
\put(139,110){.}
\put(140,109){.}
\put(141,109){.}
\put(142,108){.}
\put(143,108){.}
\put(144,107){.}
\put(145,107){.}
\put(146,106){.}
\put(147,106){.}
\put(148,105){.}
\put(149,105){.}
\put(150,104){.}
\put(151,104){.}
\put(152,104){.}
\put(153,103){.}
\put(154,103){.}
\put(155,103){.}
\put(156,102){.}
\put(157,102){.}
\put(158,102){.}
\put(159,101){.}
\put(160,101){.}
\put(161,101){.}
\put(162,100){.}
\put(163,100){.}
\put(164,100){.}
\put(165,100){.}
\put(166,99){.}
\put(167,99){.}
\put(168,99){.}
\put(169,99){.}
\put(170,99){.}
\put(171,99){.}
\put(172,98){.}
\put(173,98){.}
\put(174,98){.}
\put(175,98){.}
\put(176,98){.}
\put(177,98){.}
\put(178,98){.}
\put(179,97){.}
\put(180,97){.}
\put(181,97){.}
\put(182,97){.}
\put(183,97){.}
\put(184,97){.}
\put(185,97){.}
\put(186,97){.}
\put(187,97){.}
\put(188,97){.}
\put(189,97){.}
\put(190,96){.}
\put(191,96){.}
\put(192,96){.}
\put(193,96){.}
\put(194,96){.}
\put(195,96){.}
\put(196,96){.}
\put(197,96){.}
\put(198,96){.}
\put(199,96){.}
\put(200,96){.}
\put(201,96){.}
\put(202,96){.}
\put(203,96){.}
\put(204,96){.}
\put(205,96){.}
\put(206,96){.}
\put(207,96){.}
\put(208,96){.}
\put(209,96){.}
\put(210,96){.}
\put(211,96){.}
\put(212,96){.}
\put(213,96){.}
\put(214,96){.}
\put(215,96){.}
\put(216,96){.}
\put(217,96){.}
\put(218,96){.}
\put(219,96){.}
\put(220,96){.}
\put(221,96){.}
\put(222,96){.}
\put(223,96){.}
\put(224,96){.}
\put(225,96){.}
\put(226,96){.}
\put(227,96){.}
\put(228,96){.}
\put(229,96){.}
\put(230,96){.}
\put(231,96){.}
\put(232,96){.}
\put(233,96){.}
\put(234,96){.}
\put(235,96){.}
\put(236,96){.}
\put(237,96){.}
\put(238,96){.}
\put(239,96){.}
\put(240,96){.}
\put(241,96){.}
\put(242,96){.}
\put(243,96){.}
\put(244,96){.}
\put(245,96){.}
\put(246,96){.}
\put(247,96){.}
\put(248,96){.}
\put(249,96){.}
\put(250,96){.}
\put(251,96){.}
\put(252,96){.}
\put(253,96){.}
\put(254,96){.}
\put(255,96){.}
\put(256,96){.}
\put(257,96){.}
\put(258,96){.}
\put(259,96){.}
\put(260,96){.}
\put(261,96){.}
\put(262,96){.}
\put(263,96){.}
\put(264,96){.}
\put(265,96){.}
\put(266,96){.}
\put(267,96){.}
\put(268,96){.}
\put(269,96){.}
\put(270,96){.}
\put(271,96){.}
\put(272,96){.}
\put(273,96){.}
\put(274,96){.}
\put(275,96){.}
\put(276,96){.}
\put(277,96){.}
\put(278,96){.}
\put(279,96){.}
\put(280,96){.}
\put(281,96){.}
\put(282,96){.}
\put(283,96){.}
\put(284,96){.}
\put(285,96){.}
\put(286,96){.}
\put(287,96){.}
\put(288,96){.}
\put(289,96){.}
\put(290,96){.}
\put(291,96){.}
\put(292,96){.}
\put(293,96){.}
\put(294,96){.}
\put(295,96){.}
\put(296,96){.}
\put(297,96){.}
\put(298,96){.}
\put(299,96){.}
\put(300,96){.}
\put(301,96){.}
\put(302,96){.}
\put(303,96){.}
\put(304,96){.}
\put(305,96){.}
\put(306,96){.}
\put(307,96){.}
\put(308,96){.}
\put(309,96){.}
\put(310,96){.}
\put(311,96){.}
\put(312,96){.}
\put(313,96){.}
\put(314,96){.}
\put(315,96){.}
\put(316,96){.}
\put(317,96){.}
\put(318,96){.}
\put(319,96){.}
\put(320,96){.}
\put(321,96){.}
\put(322,96){.}
\put(323,96){.}
\put(324,96){.}
\put(325,96){.}
\put(326,96){.}
\put(327,96){.}
\put(328,96){.}
\put(329,96){.}
\put(330,96){.}
\put(331,96){.}
\put(332,96){.}
\put(333,96){.}
\put(334,96){.}
\put(335,96){.}
\put(336,96){.}
\put(337,96){.}
\put(338,96){.}
\put(339,96){.}
\put(340,96){.}
\put(341,96){.}
\put(342,96){.}
\put(343,96){.}
\put(344,96){.}
\put(345,96){.}
\put(346,96){.}
\put(347,96){.}
\put(348,96){.}
\put(349,96){.}
\put(350,96){.}
\put(351,96){.}
\put(355,96){$w=1$}
\put(51,333){.}
\put(52,317){.}
\put(53,302){.}
\put(54,288){.}
\put(55,275){.}
\put(56,262){.}
\put(57,250){.}
\put(58,239){.}
\put(59,229){.}
\put(60,219){.}
\put(61,209){.}
\put(62,200){.}
\put(63,192){.}
\put(64,184){.}
\put(65,176){.}
\put(66,169){.}
\put(67,163){.}
\put(68,156){.}
\put(69,150){.}
\put(70,145){.}
\put(71,140){.}
\put(72,135){.}
\put(73,130){.}
\put(74,126){.}
\put(75,121){.}
\put(76,117){.}
\put(77,114){.}
\put(78,110){.}
\put(79,107){.}
\put(80,104){.}
\put(81,101){.}
\put(82,98){.}
\put(83,96){.}
\put(84,93){.}
\put(85,91){.}
\put(86,89){.}
\put(87,87){.}
\put(88,85){.}
\put(89,83){.}
\put(90,81){.}
\put(91,79){.}
\put(92,78){.}
\put(93,76){.}
\put(94,75){.}
\put(95,74){.}
\put(96,73){.}
\put(97,71){.}
\put(98,70){.}
\put(99,69){.}
\put(100,68){.}
\put(101,67){.}
\put(102,67){.}
\put(103,66){.}
\put(104,65){.}
\put(105,64){.}
\put(106,64){.}
\put(107,63){.}
\put(108,62){.}
\put(109,62){.}
\put(110,61){.}
\put(111,61){.}
\put(112,60){.}
\put(113,60){.}
\put(114,59){.}
\put(115,59){.}
\put(116,59){.}
\put(117,58){.}
\put(118,58){.}
\put(119,58){.}
\put(120,57){.}
\put(121,57){.}
\put(122,57){.}
\put(123,56){.}
\put(124,56){.}
\put(125,56){.}
\put(126,56){.}
\put(127,56){.}
\put(128,55){.}
\put(129,55){.}
\put(130,55){.}
\put(131,55){.}
\put(132,55){.}
\put(133,55){.}
\put(134,54){.}
\put(135,54){.}
\put(136,54){.}
\put(137,54){.}
\put(138,54){.}
\put(139,54){.}
\put(140,54){.}
\put(141,54){.}
\put(142,53){.}
\put(143,53){.}
\put(144,53){.}
\put(145,53){.}
\put(146,53){.}
\put(147,53){.}
\put(148,53){.}
\put(149,53){.}
\put(150,53){.}
\put(151,53){.}
\put(152,53){.}
\put(153,53){.}
\put(154,53){.}
\put(155,53){.}
\put(156,53){.}
\put(157,53){.}
\put(158,53){.}
\put(159,52){.}
\put(160,52){.}
\put(161,52){.}
\put(162,52){.}
\put(163,52){.}
\put(164,52){.}
\put(165,52){.}
\put(166,52){.}
\put(167,52){.}
\put(168,52){.}
\put(169,52){.}
\put(170,52){.}
\put(171,52){.}
\put(172,52){.}
\put(173,52){.}
\put(174,52){.}
\put(175,52){.}
\put(176,52){.}
\put(177,52){.}
\put(178,52){.}
\put(179,52){.}
\put(180,52){.}
\put(181,52){.}
\put(182,52){.}
\put(183,52){.}
\put(184,52){.}
\put(185,52){.}
\put(186,52){.}
\put(187,52){.}
\put(188,52){.}
\put(189,52){.}
\put(190,52){.}
\put(191,52){.}
\put(192,52){.}
\put(193,52){.}
\put(194,52){.}
\put(195,52){.}
\put(196,52){.}
\put(197,52){.}
\put(198,52){.}
\put(199,52){.}
\put(200,52){.}
\put(201,52){.}
\put(202,52){.}
\put(203,52){.}
\put(204,52){.}
\put(205,52){.}
\put(206,52){.}
\put(207,52){.}
\put(208,52){.}
\put(209,52){.}
\put(210,52){.}
\put(211,52){.}
\put(212,52){.}
\put(213,52){.}
\put(214,52){.}
\put(215,52){.}
\put(216,52){.}
\put(217,52){.}
\put(218,52){.}
\put(219,52){.}
\put(220,52){.}
\put(221,52){.}
\put(222,52){.}
\put(223,52){.}
\put(224,52){.}
\put(225,52){.}
\put(226,52){.}
\put(227,52){.}
\put(228,52){.}
\put(229,52){.}
\put(230,52){.}
\put(231,52){.}
\put(232,52){.}
\put(233,52){.}
\put(234,52){.}
\put(235,52){.}
\put(236,52){.}
\put(237,52){.}
\put(238,52){.}
\put(239,52){.}
\put(240,52){.}
\put(241,52){.}
\put(242,52){.}
\put(243,52){.}
\put(244,52){.}
\put(245,52){.}
\put(246,52){.}
\put(247,52){.}
\put(248,52){.}
\put(249,52){.}
\put(250,52){.}
\put(251,52){.}
\put(252,52){.}
\put(253,52){.}
\put(254,52){.}
\put(255,52){.}
\put(256,52){.}
\put(257,52){.}
\put(258,52){.}
\put(259,52){.}
\put(260,52){.}
\put(261,52){.}
\put(262,52){.}
\put(263,52){.}
\put(264,52){.}
\put(265,52){.}
\put(266,52){.}
\put(267,52){.}
\put(268,52){.}
\put(269,52){.}
\put(270,52){.}
\put(271,52){.}
\put(272,52){.}
\put(273,52){.}
\put(274,52){.}
\put(275,52){.}
\put(276,52){.}
\put(277,52){.}
\put(278,52){.}
\put(279,52){.}
\put(280,52){.}
\put(281,52){.}
\put(282,52){.}
\put(283,52){.}
\put(284,52){.}
\put(285,52){.}
\put(286,52){.}
\put(287,52){.}
\put(288,52){.}
\put(289,52){.}
\put(290,52){.}
\put(291,52){.}
\put(292,52){.}
\put(293,52){.}
\put(294,52){.}
\put(295,52){.}
\put(296,52){.}
\put(297,52){.}
\put(298,52){.}
\put(299,52){.}
\put(300,52){.}
\put(301,52){.}
\put(302,52){.}
\put(303,52){.}
\put(304,52){.}
\put(305,52){.}
\put(306,52){.}
\put(307,52){.}
\put(308,52){.}
\put(309,52){.}
\put(310,52){.}
\put(311,52){.}
\put(312,52){.}
\put(313,52){.}
\put(314,52){.}
\put(315,52){.}
\put(316,52){.}
\put(317,52){.}
\put(318,52){.}
\put(319,52){.}
\put(320,52){.}
\put(321,52){.}
\put(322,52){.}
\put(323,52){.}
\put(324,52){.}
\put(325,52){.}
\put(326,52){.}
\put(327,52){.}
\put(328,52){.}
\put(329,52){.}
\put(330,52){.}
\put(331,52){.}
\put(332,52){.}
\put(333,52){.}
\put(334,52){.}
\put(335,52){.}
\put(336,52){.}
\put(337,52){.}
\put(338,52){.}
\put(339,52){.}
\put(340,52){.}
\put(341,52){.}
\put(342,52){.}
\put(343,52){.}
\put(344,52){.}
\put(345,52){.}
\put(346,52){.}
\put(347,52){.}
\put(348,52){.}
\put(349,52){.}
\put(350,52){.}
\put(351,52){.}
\put(355,52){$w=2$}
\put(10,50){\vector(1,0){350}}
\put(10,50){\vector(0,1){350}}
\put(15,400){$\theta$}
\put(15,350){$1$}
\put(50,35){0}
\put(150,35){1}
\put(390,35){$z$}
\put(150,10){$Figure \  3$ }
\end{picture}

$$ Universal \ dependence \ in \ the \ case \ of \ the \ "wide \ spectrum".$$


\begin{thebibliography}{99}
\bibitem{aero1} Kurasov V., Kinetics of aerosol formation. 1. Decay of
metastable phase on several types of heterogeneous centers (to be
published).
\end{thebibliography}
\end{document}